**Current-induced torque in a $MoS_2$-based device operating at room temperature**

Antonio Bianco [1,2,*], Michele Ceccardi [1,*], Raimondo Cecchini [3], Daniele Marré [1,2], Chanchal K. Barman [4], Fabio Bernardini [4], Alessio Filippetti [4,5], Federico Caglieris [2], Ilaria Pallecchi [2,†]

[1] Department of Physics, University of Genova, Via Dodecaneso 33, 16146 Genova, Italy
[2] CNR-SPIN, Institute for superconductors, innovative materials and devices, Corso Perrone 24, 16152 Genova, Italy
[3] CNR-ISMN, Institute for the study of nanostructured materials, via Gobetti 101, 40129 Bologna, Italy
[4] Dipartimento di Fisica, Università di Cagliari, Cittadella Universitaria, Monserrato (Ca) 09042, Italy
[5] CNR - Istituto Officina dei Materiali (IOM) Cagliari, Cittadella Universitaria, Monserrato (CA), 09042, Italy

**ABSTRACT**. Charge-to-spin and spin-to-charge conversion mechanisms in high spin-orbit materials are the new frontier of memory devices. They operate via spin-orbit torque (SOT) switching of a magnetic electrode, driven by an applied charge current. In this work, we propose a novel memory device based on the semiconducting two-dimensional centrosymmetric transition metal dichalcogenide (TMD) $MoS_2$, that operates as a SOT device in the writing process and as a non-local spin valve in the reading process. We demonstrate that stable voltage states at room temperature can be controlled by applied current and magnetic fields, owed to a locally tilted geometry and a differential voltage architecture. Ab initio calculations of orbital Hall effect (OHE) and spin Hall effect (SHE) indicate that the orbital Hall conductivity is constant and sizeable ($750\mathrm{x}(\hbar/e)\ \Omega^{-1}\ \mathrm{cm}^{-1}$) throughout the conduction gap, while the spin Hall conductivity vanishes in the gap, thus pointing to the former as the most likely responsible for the generation of the SOT in the magnetic electrode, among intrinsic mechanisms. However, also extrinsic and interface spin-orbit mechanisms likely play a role. Our device concept may be competitive in terms of energy efficiency if an acceptable trade off with stability is obtained.

[*] These two authors contributed equally to the work
[†] Contact author: ilaria.pallecchi@spin.cnr.it

## I. INTRODUCTION.

Charge-to-spin conversion mechanisms are at the basis of the emerging technology of spin-orbit torque (SOT) memory devices [1].The main mechanisms of charge-to-spin conversion are spin Hall effect (SHE) and Rashba–Edelstein effect (REE): in both cases SOT arises when a current in a nonmagnetic layer combines with the spin-orbit coupling to create a spin accumulation that exerts a torque via magnetic exchange on the magnetization of an adjacent ferromagnetic layer. In applications, SOT devices can provide better efficiency and versatility than the more established spin-transfer torque (STT) devices, presenting several advantages, such as larger switching speed, decoupling of write and read mechanisms, and larger energy efficiency [2]. SHE and REE require a large spin-orbit coupling at the band edges, and, additionally, REE requires inversion symmetry breaking and the presence of a unique polarization axis either at the bulk level or induced by some structural discontinuity. Furthermore, spin diffusion length in most materials is limited to a few tens of nm. The possibility to exploit orbital Hall effect (OHE) to operate magnetic torque devices has been only recently proposed by theoretical models [3,4, 5, 6]. Indeed, even if current-induced accumulation of orbital angular moments cannot directly exert a torque on a ferromagnet, a conversion process from orbital current to spin current within the ferromagnet does exert a torque on local magnetization. OHE does not require the breaking of either inversion or time reversal symmetry and, remarkably, can also occur in absence of spin-orbit coupling, thus hugely widening the landscape of possible implementations. On the other hand, efficient orbital to spin conversion is required for the ferromagnet.

Two-dimensional Van der Waals materials are being increasingly considered in electronics, as they present advantages like scalability down to monolayer thicknesses and atomically smooth interfaces and minimal intermixing within heterostructures. Very recently, Van der Waals materials have been employed in the fabrication of SOT devices, both in the roles of spin-orbit sources and ferromagnetic layers, introducing novel concepts and mechanisms that allowed boosting the device performances and energetic efficiency. For example, low-symmetry transition metal dichalcogenides (TMDs) enabled field-free SOT switching of out-of-plane magnetization [7,8,9] and 2D Van der Waals magnets such as $Fe_3GaTe_2$ and $Fe_3GeTe_2$ were demonstrated to be superior building blocks in SOT devices [7,10,11], thanks to their large perpendicular magnetic anisotropy and high Curie temperature.

In general, Van der Waals TMDs are a large material platform with varied properties of charge and spin transport, most of them exhibiting strong-spin orbit coupling and sizeable charge-to-spin and spin-to-charge conversion efficiency. In case of centrosymmetric semiconducting TMDs, having higher resistivities, architectures with proximitized graphene were mostly used for charge-to-spin and spin-to-charge conversion devices, where the large SOC in TMDs induces spintronic functionalities in graphene by proximity [12,13,14,15,16,17,18]. On the other hand, in devices based on lower-symmetry semimetallic TMDs such as $MoTe_2$, $WTe_2$, $TaTe_2$ and $NbSe_2$, conversion mechanisms were demonstrated in the TMDs themselves [19,20,21]. Among centrosymmetric semiconducting TMDs, hexagonal $MoS_2$ is highly stable and widely studied in literature. In $MoS_2$, where a spin-orbit coupling of 170 meV was measured by ARPES [22], efficient charge-to-spin [23,24] and spin-to-charge [25,26,27] conversion was observed up to room temperature. High-symmetry semiconducting and low-symmetry semimetallic TMDs could play different and complementary roles within SOT devices, as in the former ones spin currents are generated in the conventional configuration with mutually perpendicular spin current, charge current, and spin polarization, while in the latter ones other configurations are possible with either collinear spin current and spin polarization or with collinear spin current and charge current [9, 28, 29]. Hence, combining different TMDs in device architectures could open the way to the design and realization of non-Boolean SOT devices with multiple magnetization states. Recent calculations based on modeling predict sizeable OHE in high-symmetry TMDs [30, 31, 32, 33]. Orbital torque switching at room temperature of the van der Waals ferromagnet $Fe_3GaTe_2$ has been observed, generated by OHE in Titanium, the prototypical high-OHE material [34]. In this scenario, the possibilities of device designs with either perpendicular or planar magnetization states must be thoroughly explored, so as to take advantage of the full potential of both geometries and constituent materials, to build hybrid-anisotropy architectures [35].

In this work, we propose a new $MoS_2$-based device architecture, where the torque is exerted not on a single electrode (as in conventional SOT architectures), but simultaneously on two Co electrodes with in-plane magnetization and different shape anisotropy. In this original design the current-induced torque is exploited to switch from an equally-magnetized electrode configuration ('0' state) to a configuration where the electrodes have unaligned magnetizations ('1' state). These two, memory states are read as a differential voltage between the electrodes, and are written at room temperature by an applied current density as low as ~$3.2x10^4$ A/cm$^2$ or by a magnetic field of ~50 Oe. Such low current and field values are possible since the switch between '0' and '1' states only requires a small magnetization rotation in the shape-isotropic electrode, not the full (i.e. 180°) switch of the magnetization direction as in conventional SOT devices. On the basis of highly accurate ab-initio calculations, we suggest that the SOT may originate from large orbital-Hall current in $MoS_2$. This scenario is plausible if the orbital current is converted to spin current in the Co electrodes. On the other hand, other possible mechanisms for the SOT are extrinsic bulk SHE and interfacial REE.

## II. METHODS

### A. Experimental methods

$MoS_2$ and hexagonal BN (h-BN) flakes were obtained by mechanical exfoliation from single crystals and transferred deterministically on $Si/SiO_2$ substrates. The $MoS_2$ was 50 nm thick. Its room temperature resistivity, carrier density and mobility of the $MoS_2$ flake are $3.5x10^{-4}$ Ωm, ~$10^{18}$ cm$^{-3}$ and ~150 cm$^2$/(Vs), respectively, and the resistivity has a semiconducting temperature behaviour with thermal activation energy ~80 meV (see Supplemental Material [36]). H-BN flakes with thickness up to 10 nm were used for the fabrication of $MoS_2$/h-BN heterostack. A defective h-BN worked as resistive spacer in our device, providing a RA product of 6.4 x $10^{-5}$ Ω m$^2$ (resistance x cross section = 0.64 MΩ x 10 μm x 10 μm) (see current-voltage characteristics in the Supplemental Material [37]). Patterning of Ti(10 nm)/Au(100 nm) electrodes for current injection and Co(25nm) electrodes for spin valves were realized on the top of the heterostacks by electron beam lithography and sputtering (Ti/Au) and e-beam evaporation (Co). Measurements were carried out at room temperature, applying AC and DC currents to the $MoS_2$ flake and measuring voltage at the Cobalt electrodes. In case of superimposed AC and DC currents, $I=I_0+\Delta I \sin(\omega t)$, we always maintained $\Delta I=|I_0|$, so that the total current oscillated from zero to $I_{peak}=\pm(|I_0|+\Delta I)$ and did not change in sign during the measurement. This AC excitation allowed to reach $I_{peak}$ with halved average dissipated power as compared to applying a DC current of intensity $I_{peak}$, thus minimizing undesired heating effects. However, the signal was always measured from the DC component of the AC voltage signal and the results were not changed by the frequency of the AC excitation. An external field H up to few hundredths Oe was applied in the direction perpendicular to the current, along the axis that connects the Co electrodes. In H cycles, the sweep sequence was 0 → +100 Oe → 0 → -100 Oe → 0 → +100 Oe. In measurements at fixed H, the current was swept from $+I_{min} \rightarrow +I_{max}$ and from $-I_{min} \rightarrow -I_{max}$.

### B. Computational methods

Spin Hall conductivities (SHC) $\sigma_{SH}$ and orbital Hall conductivities (OHC) $\sigma_{OH}$ were calculated within the linear response theory by the Kubo formula [4]:

$$\sigma^{y}_{zx,SH(OH)} = \frac{e}{\hbar}\sum_n \int f_{nk}\Omega_n^{Q_y}(k)\,\frac{d^3k}{(2\pi)^3} \qquad (1)$$

$$\Omega_n^{Q_y}(k) = 2\hbar^2 \sum_{m\neq n} \mathrm{Im}\left[\frac{<u_{nk}|j_z^{Q_y}|u_{mk}><u_{mk}|v_x|u_{nk}>}{(E_{nk}-E_{mk}+i\eta)^2}\right] \qquad (2)$$

where $f_{nk}$ is the Fermi-Dirac distribution function, $< u_{nk} >$ is the periodic part of the Bloch state, whose energy eigenvalue is $E_{nk}$, $v_x$ is the x component of the velocity operator, and $j_z^{Q_y}$ is the z component of the orbital(spin) current operator $J_z^{Q_y} = (Q_y v_z + v_z Q_y)/2$, with y component of the orbital(spin) angular momentum $Q = L_y (S_y)$.

An accurate evaluation of SHC and OHC requires that the summation in Eq. (1) be performed on a very dense *k*-mesh. To avoid cumbersome ab initio calculation we made use of the Wannier interpolation technique to compute the Berry curvature term in Eq (2) on arbitrary k-points in an effortless way.

The electronic structure was obtained from ab initio calculations based on the density functional theory and using projector-augmented wave pseudopotentials [38,39] as implemented in the VASP package [40,41].

For the crystalline structure of $MoS_2$, the experimental lattice parameters $a$ = 3.1601 Å and $c$ = 12.288 Å were used, taken from ref. [42]. The first-principles calculation was performed using the generalized gradient approximation (GGA) of the Perdew, Burke, and Ernzerhof exchange-correlation functional [43]. The wave functions were expanded in a plane-wave basis with a kinetic energy cutoff of 260 eV. Reciprocal space integration was carried out using a Γ-centered 12 × 12 × 4 Monkhorst–Pack *k*-mesh.

A tight-binding Hamiltonian representation in a basis of Wannier orbitals was generated by post-processing the first-principles band structure using the WANNIER90 code [44]. We used a total of 44 orbitals for spin-up and spin-down states. As initial projections we used 20 Mo-*d* and 24 S-*p* atomic orbitals. Finally, SHC and OHC were calculated by integrating the corresponding Berry curvature over a *k*-mesh of $4 \times 10^6$ points.

## III. RESULTS AND DISCUSSION

Fig. 1 presents a sketch of the device concept and optical images of the real $MoS_2$/h-BN device with Co electrodes on top. The two Co electrodes are shaped with different aspect ratios (sizes are 5 μm x 10 μm and 10 μm x 10 μm); this makes their coercive fields different and allows the individual control of the magnetization by a planar external magnetic field and/or a longitudinal electric current in the $MoS_2$ flake. The h-BN layer serves as a resistive spacer between $MoS_2$ and Co, thus minimizing current shunting through the Co electrodes. Using finite element simulations (see Supplemental Material), we calculate that the fraction of current in $MoS_2$ shunted through Co is as low as 0.1% for the smaller electrode and 0.2% for the larger one. This asymmetry between the two Co electordes is due to their geometry and possibly concurs with shape anisotropy in enabling the differential mode operation of the device. During the cycle with the magnetic field or electric current, the differential voltage of the two electrodes ($V_{diff}$) switches between two values corresponding to state '0', that is the configuration with equally oriented magnetizations, and state '1', with magnetizations misaligned by a finite angle $\alpha$, the latter state further split into barely distinguishable sub-states for the two complementary misalignment angles $\alpha$ and $(\pi-\alpha)$.

$MoS_2$ is a non-magnetic material with both spin-current and orbital-current conversion capability: a longitudinal electric current is converted into perpendicular spin and orbital currents by spin-Hall and orbital-Hall effect, respectively, and both these may have either intrinsic or extrinsic origin, that is either related to Berry curvature or to side-jump and skew scattering events. Interface charge to spin conversion may also play a role. The spin and orbital magnetic moments carried by the Hall currents are oriented in-plane and perpendicularly to the electric current; thus, once injected into the electrodes, they can exert a torque on the magnetization of the two electrodes. Importantly, in our device the electric current in the peripheral $MoS_2$ area underneath the Co electrodes is slightly tilted off the direction perpendicular to the easy magnetization axis of the anisotropic electrode. This is a key feature to enable non collinear torques exerted by current and field, which allows switching between the the misaligned ('1' state) and the aligned ('0' state) magnetization configurations, and even current-induced switching in zero field, provided that a resetting protocol of the magnetic state is carried out by an applied field. Although our current-induced switching is not deterministic as in systems with broken symmetry [35], however we do not rule out that a local current gradient in the peripheral $MoS_2$ area underneath the Co electrodes, created by the flake edge profile that locally squeeze the current, influences the magnetic dynamics, as proposed in [45,46].

We start discussing the measurements in cycles of applied magnetic field, which allow an easier explanation of the switching mechanisms and voltage read-out; the current-driven device operation will be shown later on. Fig. 2 shows the differential voltage between the Co electrodes, measured while cycle-sweeping the applied field parallel to the direction that connects the electrodes and applying a current to the underneath $MoS_2$ (see measurement configuration in Fig. 2a). The current is composed of a DC bias $I_0$ plus and AC component of amplitude $\Delta I$, $I=I_0+\Delta I \sin(\omega t)$, with $\Delta I=|I_0|$, oscillating from zero to $I_{peak}=\pm(|I_0|+\Delta I)$, with fixed sign. Data shown in Fig. 2 are extracted as the DC component of the AC signal. In Fig. 2b, we show a cycle measured with current $|I_{peak}|$=5 mA, representative of all cycles measured with $|I_{peak}|$ exceeding a threshold ~1.6 mA. A clear butterfly-shaped hysteresis is seen, however, the abrupt jumps occur at ±5 Oe and ±50 Oe as the field is on its way back to zero, at odds with the usual behaviour of spin valves, where jumps should occur in the branches of increasing field magnitude. Indeed, as indicated by the arrows, the butterfly-shaped hysteresis traces a clockwise cycle for positive H and an anticlockwise cycle for negative H.

The magnitude of the voltage jump is asymmetric with respect to the sign of the applied field, and this asymmetry is swapped when inverting the sign of the applied current. On the other hand, the jump magnitude does not have a clear dependence on the magnitude of the current (see other cycles in Supplemental Material), as long as it exceeds the above mentioned threshold intensity (we will come back to this result while describing our tentative micromagnetic model for the current-induced writing process). This latter observation rules out any simple ohmic mechanism as responsible for the jumps and also departs from a squared current dependence expected for Joule heating.

In Fig. 2c, we show a comparison between a cycle measured with $I_{peak}$ exceeding the threshold ~1.6 mA and a cycle measured with current $|I_{peak}|$=0.6 mA, that is below the threshold. Ascending and descending |H| branches are displayed in different panels. In the case of the cycle measured with below-threshold current, a typical spin-valve hysteresis is observed, with voltage jumps in the branches of increasing field magnitude at coercive fields ±7 Oe and ±54 Oe.

We consider three possible origins for the $V_{diff}$ dependence on the relative magnetization alignment in the electrodes. As a first hypothesis, it could originate from anisotropic magnetoresistance (AMR) of Co electrodes, where a fraction of current applied to $MoS_2$ is shunted. AMR in our polycrystalline Co reaches ~0.4% for 90° rotation of magnetization with respect to the direction of the current (see Supplemental Material), in agreement with literature [47]. Considering (i) that only 0.1-0.2% of the applied current is shunted through Co thanks to the resistive h-BN spacer; (2) that our voltage signal is likely generated by a rotation of magnetization much smaller than 90°; (3) that our voltage signal increases/decreases by 10% - 25 % at switching in different measurements, we conclude that AMR is unlikely responsible for the hysteretic jumps of the $V_{diff}$ voltage signal. Furthermore, in current sweeps (Fig.s 3 and 4 and Fig. S3 of the Supplemental Material), we observe that the jump between two different states appears as a rigid shift of ohmic current-voltage characteristics, i.e. with a change of intercept and unchanged slope. On the contrary, a magnetic switching and related change in AMR signal would be seen as a change of both intercept and slope.
The second hypothesis for the origin of $V_{diff}$ is anomalous Hall effect in Co, driven by a perpendicular thermal gradient related to Joule heating. Also this hypothesis must be ruled out, as the voltage is picked up between separate electrodes.
The third possible origin of $V_{diff}$ is the underlying spin and orbital moment accumulation at the $MoS_2$ interface, which is detected by the Co electrodes as spin-sensitive voltage probes [48]. In this scenario, $V_{diff}$ is a signature of charge-to-spin and/or charge-to-orbital conversion in $MoS_2$. In absence of this effect, $V_{diff}$ would not be altered since any direct spin flow between the electrodes and across $MoS_2$ is ruled out by the distance of the electrodes (tens μm), which by far exceeds any realistic estimation of the spin-diffusion length in $MoS_2$.

Considering the switching process in H cycles, while the spin-valve hysteresis measured with sub-threshold currents, with jumps in the ascending |H| branches, is driven by the applied field, the hysteresis measured with above-threshold currents, with jumps in the descending |H| branches, must be necessarily driven by the applied current. In some studies [49,50], it is suggested that inverted hysteresis is possible due to magnetic coupling between two materials with unequal magnetic properties, one exhibiting higher and softer magnetic moments and the other lower and harder moments, as in such systems, one of two magnetizations might be reversed with the help of an external driving force (spin-orbit torque) during the removal of a magnetic field to lower the total energy of the system. In our system, these materials are likely represented by $MoS_2$ with accumulated magnetic moments and Co. The comparison of the coercive fields observed in cycles measured with above-threshold current (±5 Oe and ±50 Oe) and with below-threshold current (±7 Oe and ±54 Oe) indicates that the efficiency of the equivalent in-plane field of the applied current is ~$3x10^{-5}$ Oe/(A $cm^2$).

We now present in Fig. 3 the current cycles. In Fig. 2, the current was kept fixed and H was cycled, so that a hysteresis as a function of H was evidenced. On the contrary, in Fig.s 3a and b, H is kept fixed and the current is swept according to a specific protocol, so that a hysteresis as a function of the current is evidenced. Specifically, a positive H was fixed a H=+100 Oe and the current was swept twice from $I_{peak}$=+1 mA ($\Delta I=|I_0|$, $I_0$=+0.5 mA) to a maximum positive value $I_{peak}$=+6mA ($\Delta I=|I_0|$, $I_0$=+3mA) in 1 mA steps, and then swept twice from $I_{peak}$=-1mA ($\Delta I=|I_0|$, $I_0$=-0.5mA) to a negative value $I_{peak}$=-6mA ($\Delta I=|I_0|$, $I_0$=-3mA), in 1mA steps. The same procedure was successively applied at fixed H=-100 Oe, H=+25 Oe, H=-25 Oe (Fig. 3b) and finally in zero applied field (Fig. 3a), in this sequence. As indicated by the arrows that trace the sequence of data points, in each pair of identical current sweeps there is an offset of the second curve with respect to the first curve, for both negative and positive current, as well as for both negative, positive and zero field. This offset indicates that an additional non-ohmic voltage is present in some curves. Remarkably, the cycle is repeated identical indefinitely and deterministically. Note that any additional current sweep with the same current polarity after the second sweep results in no further switching, confirming the role of the current direction in the process. The plots of the intercepts obtained from linear fits to data of each current sweep are shown in Fig. 3c. Two features are clearly observed: a two level behavior and a hysteresis, whose clockwise/anticlockwise direction is determined by the sign of the current and the applied field. This hysteresis represents a deterministic switching between two levels controlled by the applied current and its sign. Most remarkably, current-induced switching occurs even in zero field (Fig. 3a), even if the relation between '0' to '1' or '1' to '0' switching and the sign of the applied current depends on the initial magnetic configuration of the electrodes. It can be also noted that, for a given initial magnetic configuration of the electrodes, the hysteresis loops at zero field and at H=±25 Oe are all cycled in the same direction, as indicated by the arrows. This means that zero or low fields (< 50 Oe) do not influence the irreversible current-induced switching, which is only determined by the sign of the applied current, as long as it is above a threshold value. On the contrary, at fields H=±100 Oe - that is for fields larger than the switching field H= 50 Oe identified in the hysteresis H cycles of Fig. 2 - the direction of the cycles is inverted for positive and negative field, indicating that the irreversible current-induced switching is determined not only by the sign of the applied current, but also by the sign of the applied field. Further switching processes obtained with different protocols of applied current sweeps and fields are shown in the Supplemental Material.
In order to determine this switching threshold value of the current, we performed current sweeps in zero field, reaching increasingly higher currents $I_{peak}^{max}$ in each successive sweep, until a switch is observed, in the form of an offset voltage. In Fig. 4, we observe that as the current $I_{peak}$ exceeds 1.6 mA in zero field, the sets of sweep data points are shifted by an offset. We thus identify $I_{peak}$=1.6 mA as the switching current, corresponding to switching current density $1.6x10^4$ A/cm$^2$. On the other hand, non-perturbative reading of the state can be done with sub-threshold currents, provided that the signal to noise ratio allows to appreciate the ~1μV difference between the two voltages states. We noted that reading currents well below the threshold (e.g. $I_{peak}$=0.4 mA or smaller) are recommended, as approaching the threshold current (1 mA ≤ $I_{peak}$ ≤ 1.4 mA) some switching instability occasionally occurs, due to partial domain switching. This is seen both in Fig. 4, where data points at intermediate currents are somewhat scattered off a straight ohmic line and in Fig.s 3a and b, where an abrupt change in current sign to ±1 mA switches the system to the opposite state.

To understand the origin of the mechanisms occurring in the $MoS_2$ layer of our device, we performed ab-initio calculations of spin-Hall conductivity (SHC) $\sigma_{SH}$ and orbital-Hall conductivity (OHC) $\sigma_{OH}$ in bulk $MoS_2$. Our theoretical analysis focuses exclusively on the intrinsic contributions, which originate from the Berry curvature of the bulk band structure, even if extrinsic contributions cannot be ruled out. In coherence with the experimental setting, we are specifically interested in the spin and/or orbital current flowing perpendicularly to the Co/h-BN/$MoS_2$ interface (out-of-plane direction), with angular momentum polarization parallel to the interface (in-plane direction). This is represented in Fig.5 by the coefficient $\sigma_{zx,SH(OH)}^{y}$ of the conductivity tensor as a function of the Fermi energy, where $x$ is the in-plane direction of the applied electric field, $z$ the spin or orbital current direction, and $y$ the spin or orbital polarization direction. The conductivities remain independent on the electric field direction, as long as it lies in-plane. Therefore, the $\sigma_{zy,SH(OH)}^{x}$ coefficient of the conductivity has the same absolute value of $\sigma_{zx,SH(OH)}^{y}$, but opposite sign. Two striking features are immediately apparent in the figure: first, the OHC values are huge and orders-of-magnitude larger than their spin counterpart in all the examined energy range. Second, while the SHC exactly vanishes in the fundamental band gap, as expected for a non-topological insulator, the OHC does not; in fact, it displays a constant plateau, with a remarkably large value of about 750x($\hbar$/e) $\Omega^{-1}$ cm$^{-1}$ throughout the conduction gap. This outstanding feature has been already noticed and analyzed in previous works based on tight-binding models [30, 31] and attributed to the peculiar orbital texture in $k$-space around the valley points $K$ and $K$' of the Brillouin zone. Resistivity measurements (see Supplemental Material) show that our $MoS_2$ sample is a non-degenerate n-type semiconductor, thus we expect that the reference Fermi energy falls within the upper part of the $MoS_2$ conduction bandgap, where SHC is absent.
This result points to a scenario where, at least at the intrinsic level, orbital current could be the driving force behind the differential voltage observed in the device. We remark, however, that the SHC is particularly small for the out-of-plane spin current orientation, while for planar directions (see a complete account in the Supplemental Material) it is sizeable in the vicinity of band extrema, albeit nowhere near the OHC values. Thus, we conclude that while in other $MoS_2$-based architectures a thermally activated charge current could produce spin-charge conversion in $MoS_2$, the orbital-charge conversion will still be the dominant mechanism, anyway.

As extensively studied in literature, in ferromagnets the orbital current must be converted into a spin current in order to exert a torque. Spin-orbit conversion efficiency of ferromagnetic Co is reported by ab-initio calculations to be fairly large [51] and, depending on the system, torques generated by bulk orbital currents can be even larger in Co than in Ni [52]. Combined ab initio and spin dynamics calculations indicate that the presence of induced orbital moments contribute significantly to the magnetization switching of Co in Pt/Co bilayers [53]. Experiments reported enhanced torques in Co/Pt heterostructures, by adding an interlayer of Al, a light metal responsible for a large Orbital Rashba-Edelstein effect at the Co/Al interface [54]. However, spin-orbit conversion efficiency in Co is not as large as in Nickel and this could indicate that a spin current originating from extrinsic SHE or REE in $MoS_2$, rather than an orbital current originating from extrinsic OHE in $MoS_2$, may be responsible for the torque in Co.

As for the mechanisms occurring in the Co layer of our device, we do not have a direct probe of magnetic dynamics, hence we propose a tentative micromagnetic model for the writing process, responsible for the hysteretic voltage steps observed in the experiments of H cycles with above-threshold currents (Fig. 2b). Whereas in H-driven magnetization switches of spin valves the voltage steps occur in the branches of H cycles where H is increasing in magnitude, in H cycles with above-threshold currents the voltage steps appear in the branches of decreasing H magnitude, as H is swept back to zero (see the direct comparison in Fig. 2c). This indicates that not just a H-driven magnetization switch takes place. We argue that, instead, a current-assisted switch takes place, namely a SOT process.
Fig. 6 depicts a top-view sketch of microscopic mechanisms, according to our interpretative picture. Three main factors come into play:
- non collinear directions of injected magnetic moments and Co magnetization, determined by the irregular shape of the $MoS_2$ flake and the peripheral position of the Co electrodes;
- polydomain nature of Co, with domain size that increases with applied field and/or applied current;
- spin-orbit torque and reversible rotation of smaller Co domains or irreversible switching of larger Co domains.
More specifically, according to our model, in the first branch of the H cycle, the fragmented domains grow larger with increasing H, but still too small and thus reversibly rotated between angles from 0 to $\alpha$ at the AC frequency; the differential voltage between the Co electrodes is in its collinear level, '0' state, with Co electrodes having the same average magnetization direction. In the second branch, as H decreases, the domains have grown larger, hence they have larger coercivity. When the torque due to the field drops below coercivity, the enlarged domains are irreversibly switched along the direction of the injected magnetic moments, i.e. by an angle $\alpha$. As the electrodes have different shape anisotropy, these switching fields are different for the two electrodes, being 50 Oe and 5 Oe, respectively. When H is within these two values, the differential voltage between the Co electrodes is in its misaligned level, '1' state, with Co electrodes having the non collinear magnetization directions. The third and fourth branches behave similarly, with magnetization of Co electrodes changed by $\pi$ and the misalignment angle changed to ($\pi$-$\alpha$). For inverted sign of the applied current, the roles of H-positive and H-negative branches get swapped.
Our micromagnetic model sketched in Fig. 6, applies not only to the H cycles, but also to the current cycles (Fig.s 3 and 4). We assume that not only the increasing field, but also the increasing current drives the enlargement of domains in the first current sweep, so that while in the first current sweep the domains rotate reversibly under the opposed current- and field-induced torques, in the successive identical current sweeps the domains switch irreversibly to the direction of the injected magnetic moments. As the current is reversed to a value of opposite sign that is close to or above the threshold value, the domains get fragmented again and a new current sweep that makes the domains grow is necessary to enable the irreversible switching in the successive sweep. As it comes out from the summary of switching sequences displayed in Fig. 3c, in this regime of current-driven domain growth, zero or low fields ($H < 50$ Oe) do not influence the irreversible current-induced switching, which is only determined by the sign of the applied current. On the other hand, at higher fields $H > 50$ Oe, the irreversible current-induced switching is determined by both the sign of the applied current and the sign of the field. In order to better explain the deterministic character of the switching in zero field, in Fig. 7 we sketch our model for the evolution of magnetic domains in the Co electrodes in a current cycle as that of Fig. 3a.
In the in the Supplemental Material we explain in more detail our proposed micromagnetic description of current-induced switching in H cycles and in current sweeps.

It is necessary to discuss some possible alternative or concurring effects that may generate torques on the local Co magnetization or play a role in the switching process.
1) Thermal effect may concur with torque in triggering magnetic switching. However, the magnitude of measured voltage steps does not depend on the current magnitude, as long as it is above threshold, as opposed to a quadratic dependence on the applied current expected in presence of dominant Joule heating. Moreover, the sample temperature was measured throughout all the measurements for each data point and it was observed to vary within one tenth of K. Most of all, Joule heating may at most assist magnetic switching, but it can account neither for the observed hysteresis, nor for the observed dependence of the effects on the current sign. Hence, we can claim that Joule heating effect, if any, is minor.

2) Oersted fields may determine switching of in-plane magnetization in torque devices. In our case, we estimate the Oersted field generated in proximity (distance << sample width) of a sample of thickness 50 nm with uniform current density $3x10^4$ A/cm$^2$ to be as low as H=0.1 Oe, thus it should play a negligible role in the magnetization switching.
3) Self-torque in Co electrodes can be considered. We carried out finite element simulations on our device geometry (see Supplemental Material). Thanks to the h-BN interlayer resistive spacer between $MoS_2$ and Co, which inhibits current shunting, the current flowing in the Co electrodes is 2 orders of magnitude smaller than the current flowing in the $MoS_2$ layer. Considering the values of applied current in our work, we estimate that current density in the Co electrodes reaches at most 100-200 A/cm$^2$, certainly too low a value to trigger self-torque switching of local magnetization. Indeed, our second harmonic Hall measurements in Co Hall bars confirm that no self-torque is observable, even for applied current densities up to $10^5$ A/cm$^2$ (see Supplemental Material).

We remark that our architecture radically differs from the perpendicular-magnetization SOT devices, where the magnetization switching by spin torque requires an additional strategy capable to break the mirror symmetry (e.g. chiral symmetry breaking, exchange bias, interlayer exchange coupling, spin reflexion canted antiferromagnet interlayer [35,55,56]). In our device the only relevant parameter which modulates the differential voltage is the mutual misalignment angle $\alpha$ (or $\pi$-$\alpha$), that is much smaller than the 180 degrees switching of conventional SOT devices. As a consequence, the two magnetization states can be switched back and forth by a switching current $I_{peak}$~1.6 mA, corresponding to a switching current density ~$3.2x10^4$ A/cm$^2$. A direct comparison with ~$10^6$ A/cm$^2$ values measured in SOT devices based on TMDs and topological insulators [1,2,10,11,57,58,59,60] and ~$10^5$ A/cm$^2$ measured in a SOT device based on $TaIrTe_4$ [61] would suggest high energy efficiency, however stability issue should be also considered for electronic applications. State stability is certainly poorer in our device, due to the low magnetic anisotropy of Co electrodes and the low misalignment angle $\alpha$ between the two states.
In our device switching between voltage states is deterministic in zero field, once an initialization magnetization protocol is used, with application of above-threshold field and current applied, to fix a given initial configuration of magnetizations, so that the polarity of the current that must be applied to switch the state is established. As explained in the model described in the Supplemental Material, this occurs because different micromagnetic configurations corresponds to the '0' state (polydomain with the same average magnetization direction of the two Co electrodes or monodomain with the same magnetization direction) and likewise different micromagnetic configurations corresponds to the '1' state ($\alpha$ or ($\pi$-$\alpha$) misalignment angle between the magnetization directions of the two Co electrodes). On the other hand, it is possible that a current density gradient is created locally by the tilted flake edge, so that gradient-induced deterministic torques are generated.
Examples of sequences of deterministic switching events are shown in Fig.s 3 and 4, and in the Supplemental Material.

For what concerns future strategies of device optimization, we can envisage several areas of improvement : a) achieve a better control of the tilting angle $\alpha$ by lithographic patterning of the flake; b) better control on the resistance of the spacer in terms of thickness or material; c) choice of ferromagnetic electrodes with larger spin-orbit coupling (e.g. Ni in place of Co) to maximize the orbital to spin moment conversion [6]; d) optimization of the dissipated power density (~$5x10^3$ W/cm$^2$) to meet the technological power density target of 10 W/cm$^2$; e) shaping magnetic electrodes to improve magnetic anisotropy and achieve better stability of the memory state for sub-threshold currents; f) exploration of other two-dimensional materials as generators of spin and/or orbital currents.

## V. CONCLUSIONS

In this work we present a novel spintronic device capable of current-induced switching at room temperature, using writing current densities as low as ~$3.2x10^4$ A/cm$^2$. The switch between two fairly stable in-plane magnetic states is achieved through an original design that combines non-local spin-valve and spin-torque concepts for reading and writing operations, respectively. Two aspects characterize most effectively our design: (ii) the differential character of the output voltage measured between two magnetic electrodes with different shape anisotropy, and (ii) the local tilting of the current density induced by the profile of the flake edge, which, combined with the Co magnetic easy axis and the applied field, allows switching between aligned and misaligned electrode magnetization states.

Concerning the fundamental interpretation of the observed phenomenon, our ab-initio results indicate that, among intrinsic effects, the vertical Hall current generated in $MoS_2$, which accumulates at the interface, is, in fact, not a spin-Hall current, but an orbital-Hall current. We emphasize that for a semiconductor like $MoS_2$, only the orbital-Hall conductivity can be sizeable in the band gap, since the spin-Hall conductivity must vanish, and this is a fundamental feature, not a matter of specific calculation. It follows that only the orbital-Hall conductivity can contribute, at least for what concerns non-degenerate doping regimes. In this microscopic scenario a strong orbital current is injected into the Co electrode across the Co/h-BN/$MoS_2$ interface, with an in-plane polarization perpendicular to the electric field direction. The spin-orbit coupling of the Co layer converts the injected orbital moments into spin moments and exerts a torque on the electrode magnetizations. Therefore, this microscopic phenomenon can be described as an orbital injection-induced torque. The low spin-orbit conversion efficiency of Co suggests that additional effects in $MoS_2$ (e.g. extrinsic spin-Hall, interface Rashba-Edelstein) could also come into play.

The constituent materials are worth a final consideration. Our device is based on the semiconducting Van der Waals $MoS_2$ dichalcogenide, whose OHE is predicted by our theoretical calculations to be sizeable. The family of Van der Waals materials includes countless high spin-orbit dichalcogenides, in which charge-to-spin conversion mechanisms generate spin currents with both out-of-plane and in-plane spin polarizations. Moreover, many two-dimensional Van der Waals ferromagnetic compounds exist, such as $Fe_3GeTe_2$, $Fe_3GaTe_2$, $CrI_3$ and $CrT_2$, with either in-plane or out-of-plane anisotropy and large Curie temperatures, even up to room temperature. From these premises and considering the advancements in their fabrication technology, the platform of Van der Waals materials is ideal for developing hybrid-anisotropy device architectures [35] that rely on the complementarity of in-plane anisotropy systems for high-speed operation and out-of-plane systems for ultra-high-density nonvolatile storage.

## ACKNOWLEDGMENTS

The work received funding from Italian Ministry of University and Research MUR, financed by the European Union - Next Generation EU, within the PRIN 2022 call, project SUBLI “Sustainable spin generators based on Van der Waals dichalcogenides” contract n. 2022M3WXE7. I.P. and A.F. also acknowledge the “Network 4 Energy Sustainable Transition–NEST” project, award number PE0000021, funded under the National Recovery and Resilience Plan (NRRP), Mission 4, Component 2, Investment 1.3 - Call for tender No. 1561 of 11.10.2022 of Italian Ministero dell'Università e della Ricerca (MUR); funded by the European Union–NextGenerationEU. A.F.also acknowledges project PRIN 2022 TOTEM, grant n. F53D23001080006, funded by Italian Ministry of University and Research (MUR), project PNRR-PRIN 2022 MAGIC, grant n. F53D23008340001, funded by EU. The authors acknowledge helpful contribution of Michele Bellettato for the clean room processing of the samples.

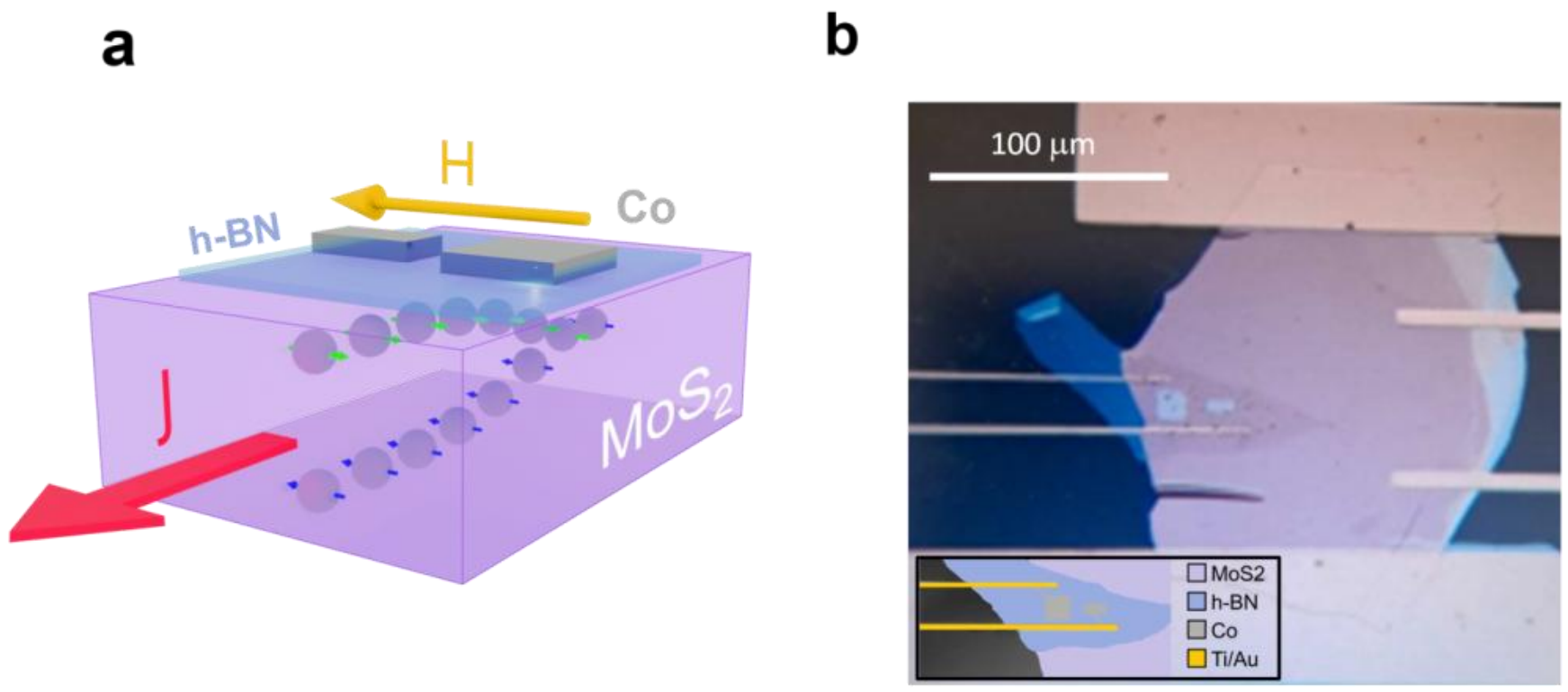

**Figure 1:** a) sketch of the device concept, evidencing the direction of the applied charge current (red arrow), applied magnetic field (yellow arrow) and separation of electrons with opposite magnetic moments inside the $MoS_2$ flake (spheres with green and blue arrows). b) Optical image of the device, showing $MoS_2$ and h-BN stacked flakes, contacting Ti/Au stripes and Co pads with different aspect ratio for control of the respective coercive fields. In the inset, a sketch indicates the constituent materials of the device.

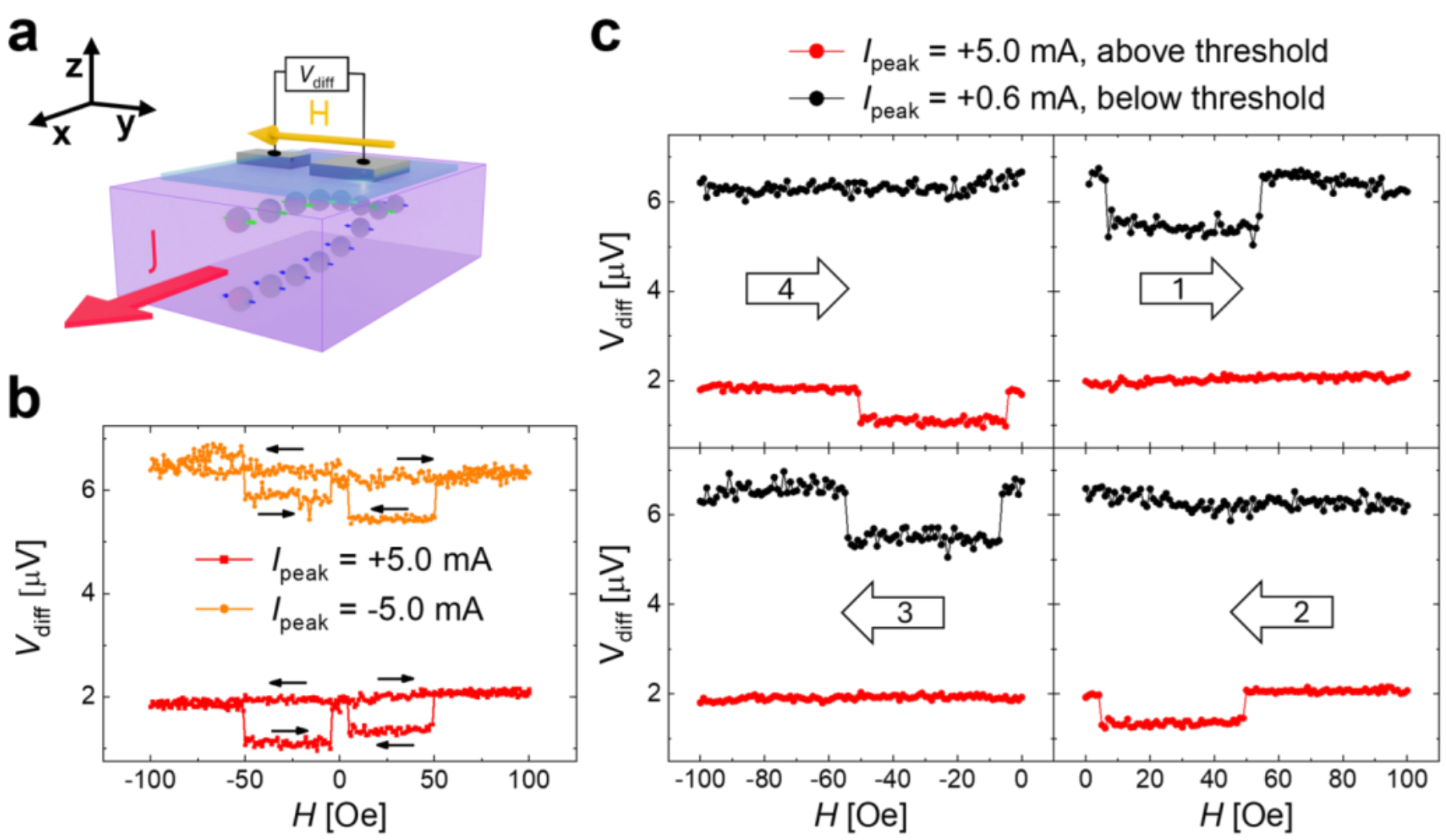

**Figure 2:** a) sketch of the configuration for the measurement differential voltage between the Co electrodes. b) H cycles of differential voltage between the Co electrodes, for applied charge currents in $MoS_2$ $I=I_0+\Delta I \sin(\omega t)$, with $I_{peak}=2xI_0=+5mA$ and -5mA, and $\Delta I=5mA$. The arrows indicate the H sweep direction, evidencing steps in the decreasing |H| branches. These cycles are representative of all cycles measured with currents $|I_{peak}|>1.6$ mA. c) First to fourth branches of H cycles measured $I_{peak}=+5mA$, above-threshold current, and $I_{peak}=+0.6mA$, below-threshold current. The arrows indicate the H sweep direction, with voltage steps only in the decreasing |H| branches, second and fourth branches, for $I_{peak}$ above threshold and voltage steps only in the increasing |H| branches, first and third branches, for $I_{peak}$ below threshold.

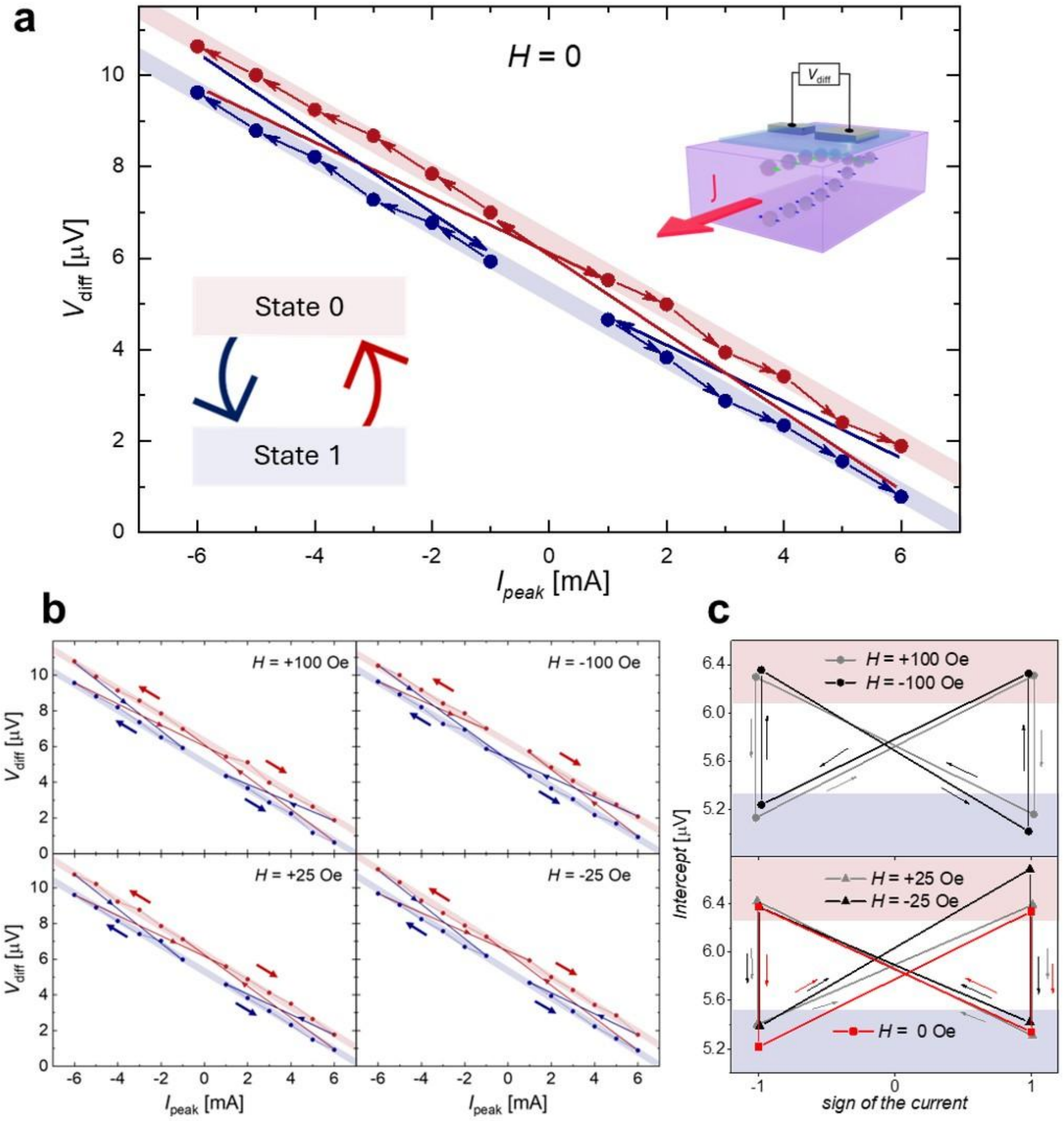

**Figure 3:** Differential voltage between Co electrodes, measured in sweeps of applied current $I=I_0+\Delta I \sin(\omega t)$, with positive current ($\Delta I=|I_0|$, $I_{peak}=2xI_0$) for two successive sweeps and negative current ($\Delta I=|I_0||$, $I_{peak}=2xI_0$) for other two successive sweeps in zero field (panel a) and at fixed magnetic field (H=+100 Oe, H=-100 Oe, H=+25 Oe, H=-25 Oe (panel b). The arrows indicate the sequential order of data points. The intercepts of the linear fit of sweeps are plotted as a function of the sign of the applied current (panel c). In all panels, the backdrop shadings are a guide to the eye to identify the two states.

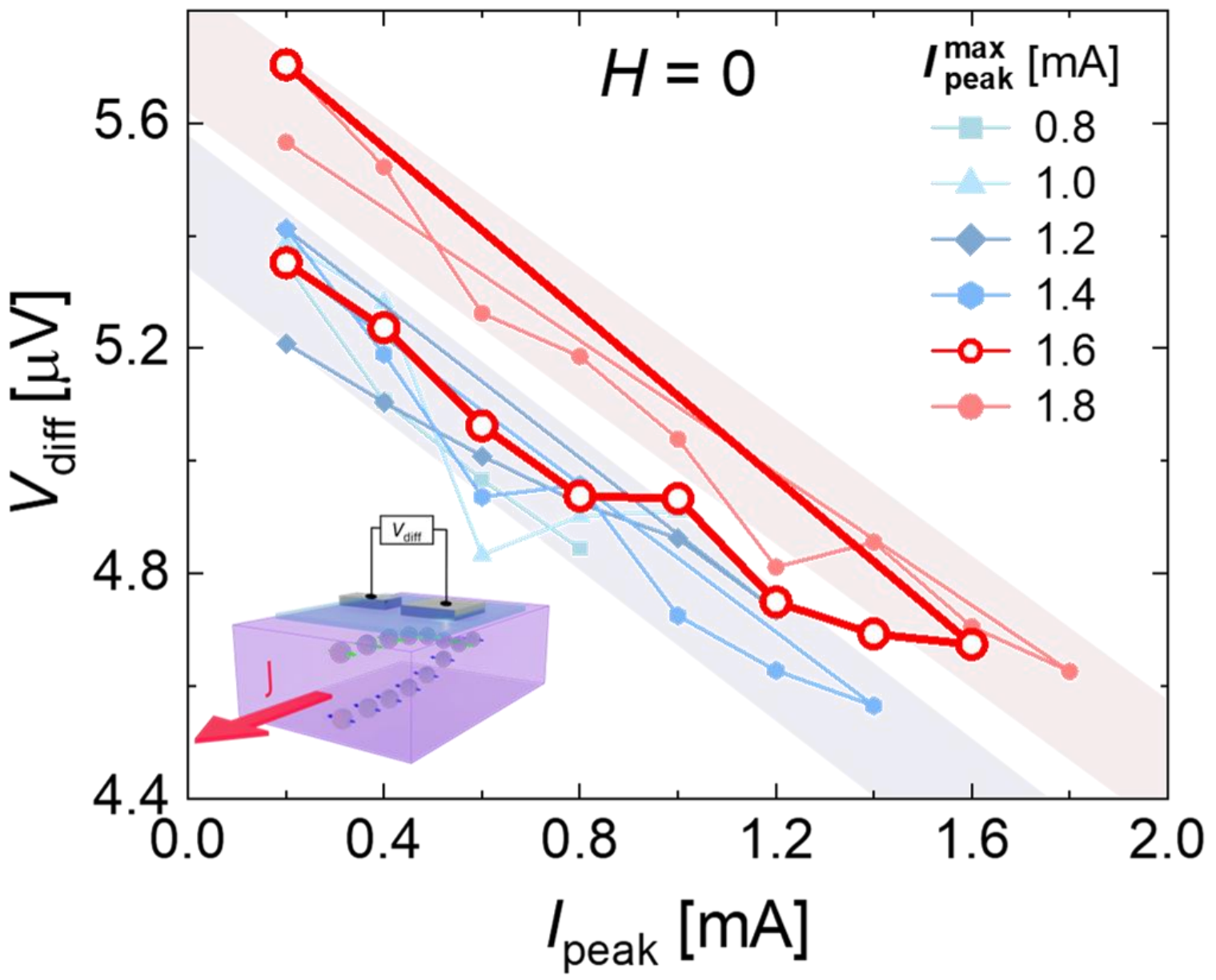

**Figure 4:** Differential voltage between Co electrodes, measured in successive sweeps of applied current $I=I_0+\Delta I \sin(\omega t)$, up to an increasing maximum intensity $I_{peak}^{max}$, in zero field.

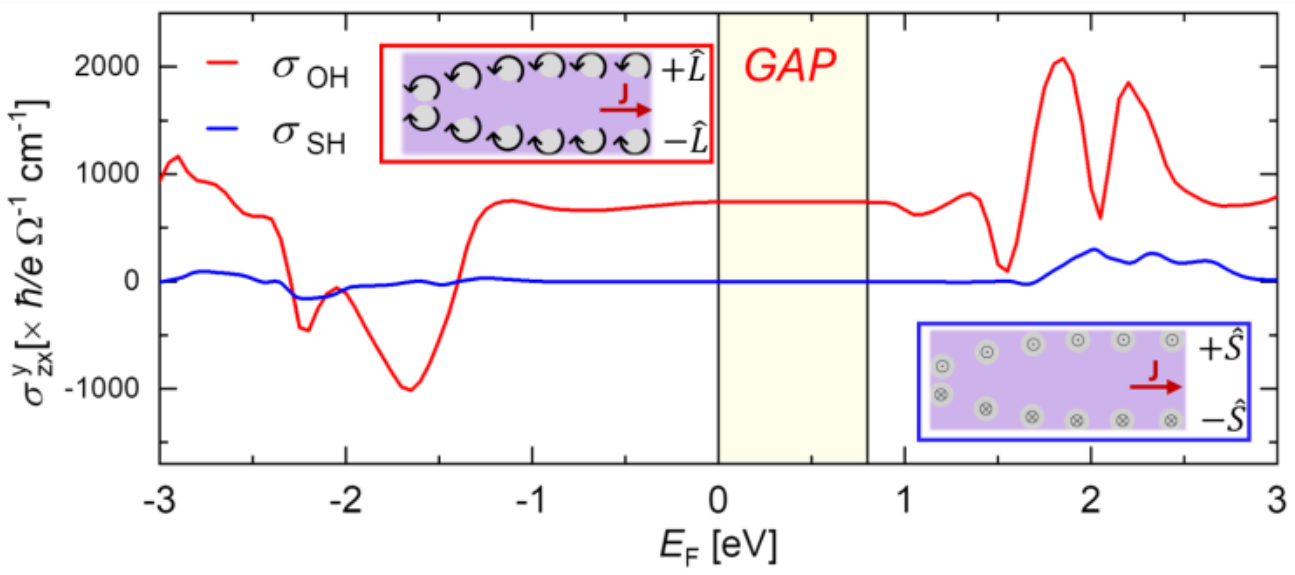

**Figure 5:** Calculated spin Hall conductivity $\sigma_{SH}$ (blue) and orbital Hall conductivity $\sigma_{OH}$ (red) of hexagonal $MoS_2$. The spin / orbital current flows parallel to the c-axis (z-direction) of the $MoS_2$ crystal. The electric field (x-direction) and the spin/orbital polarizations (y-direction) are orthogonal each other, both lying in the a-b crystal plane. The sketches in the insets depict orbital (upper left) and spin (lower right) Hall effects.

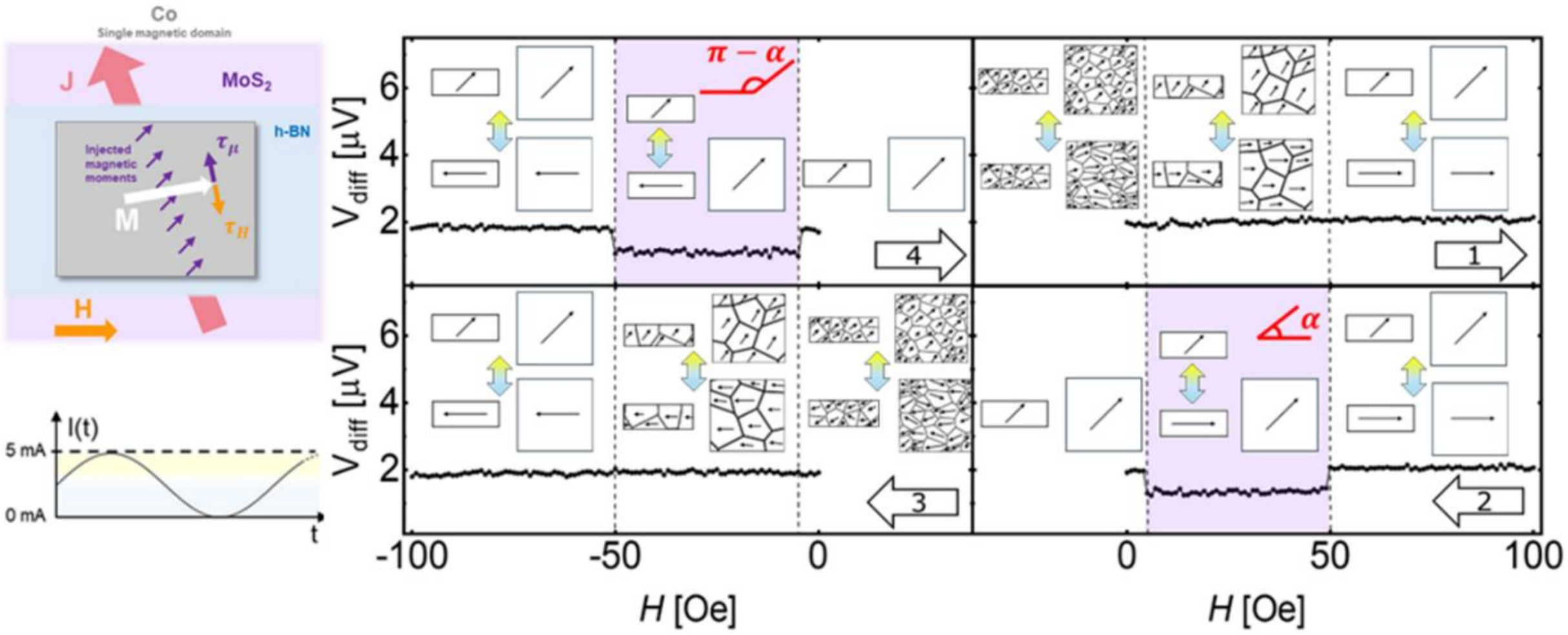

**Figure 6:** Main panel: sketch of our tentative model for magnetic domains in the Co electrodes, as they evolve throughout a H cycle. The four panels depict the four branches of the H cycle. In each panel, a H cycle measured with above-threshold current helps correlating the evolution of the domains with the measured signal along the cycle. Upper left panel: example of directions of applied current, magnetization of a domain, injected magnetic moments and related anti-damping torque ($\tau_\mu$) on the local magnetization, applied field and related torque ($\tau_H$) on the local magnetization. Lower left panel: sinusoidal waveform of the applied current. The yellow/cyan shadow in the sinusoidal waveform corresponds to the same shadow of the double-edged arrows in the main panel, indicating the regimes where the magnetization rotates reversibly between angles from 0 to $\alpha$ at the AC frequency of the applied current.

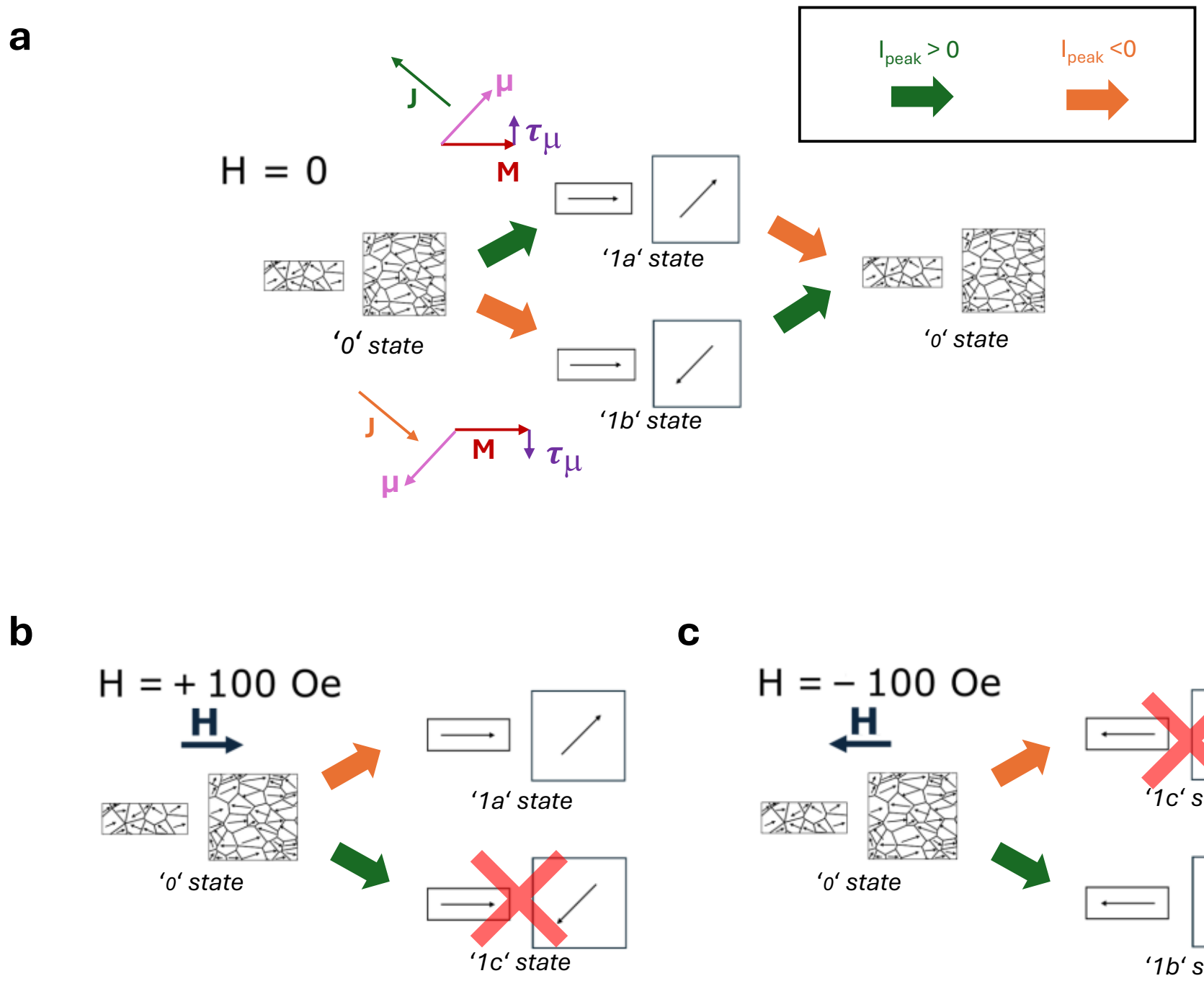

**Figure 7:** Sketch of our tentative model for magnetic domains in the Co electrodes, as they evolve throughout a current sweep in zero field (upper panel) and in a magnetic field applied field along the direction of the smaller Co electrode easy axis exceeding the coercive field (lower panel). Note that while ‘1a’ and ‘1b’ states can be realized, the ‘1c’ state is unlikely to be realized, due to the energetically unfavorable orientation of the larger Co electrode magnetization with respect to the applied field, and this explains why in presence of a magnetic field larger than the coercive field, the direction of the current that switches from ‘0’ state to ‘1’ state is opposite for opposite directions of the field, as observed experimentally (see Fig. 3c). Directions of applied current, applied field, magnetization of domains M, injected magnetic moments μ and related torque ($\tau_\mu$) are indicated, while the upper right inset is a color legend for the direction of the applied current.

**Supplemental Material**

**Highly efficient field-free switching by current-induced torque in a $MoS_2$-based device operating at room temperature**

Antonio Bianco [1,2,**], Michele Ceccardi [1,**], Raimondo Cecchini [3], Daniele Marré [1,2], Chanchal K. Barman [4], Fabio Bernardini [4], Alessio Filippetti [4,5], Federico Caglieris [2], Ilaria Pallecchi [2,*]

[1] *Department of Physics, University of Genova, Via Dodecaneso 33, 16146 Genova, Italy*
[2] *CNR-SPIN, Institute for superconductors, innovative materials and devices, Corso Perrone 24, 16152 Genova, Italy*
[3] *CNR-ISMN, Institute for the study of nanostructured materials, via Gobetti 101, 40129 Bologna, Italy*
[4] *Dipartimento di Fisica, Università di Cagliari, Cittadella Universitaria, Monserrato (Ca) 09042, Italy*
[5] *CNR - Istituto Officina dei Materiali (IOM) Cagliari, Cittadella Universitaria, Monserrato (CA), 09042, Italy*
[**] *These two authors contributed equally to the work*
[*] *Corresponding author: Ilaria.pallecchi@spin.cnr.it*

# 1. H cycles at different AC above-threshold currents and frequencies

Fig.s S1 and S2 show the differential voltage between the Co electrodes, measured while cycle-sweeping the applied field parallel to the direction that connects the electrodes and applying a current to the underneath $MoS_2$. Different values of above-threshold current (Fig. S1) and different frequencies (Fig. S2) are displayed. In Fig. S1, the magnitude of the voltage steps exhibits no clear dependence on the magnitude of the current, as long as the latter exceeds the threshold $|I_{peak}|>1.6$ mA. Only at the largest current 6 mA the step is slightly smaller, possibly due to slight Joule heating and consequent thermally assisted magnetic switching. This effect is however minor, thanks to the large Curie temperature of Co.
In Fig. S2, it is shown that measurement frequency plays no significant role either.

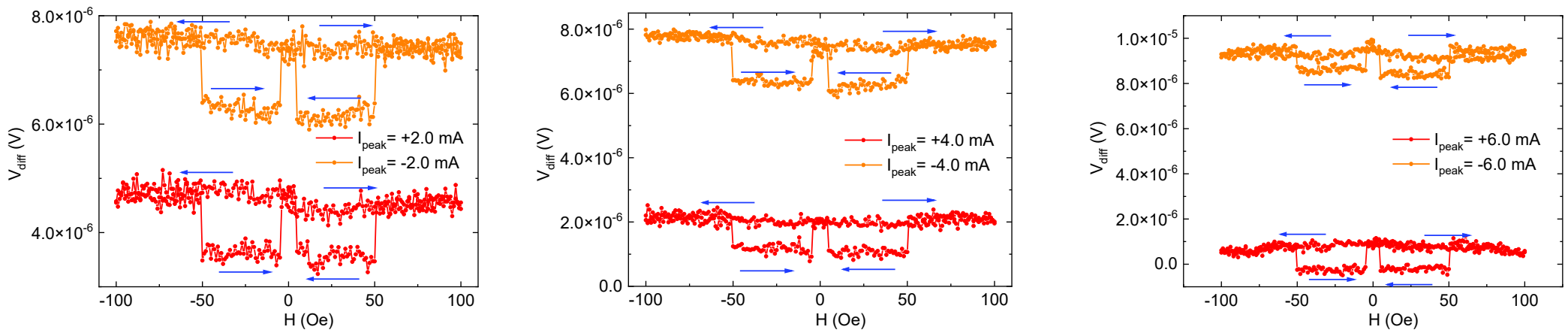

**Fig. S1.** H cycles of differential voltage between the Co electrodes, for applied charge currents in $MoS_2$ $I=I_0+\Delta I \sin(\omega t)$, with positive and negative above-threshold values of $I_{peak}=2xI_0$ and $\Delta I=|I_0|$.

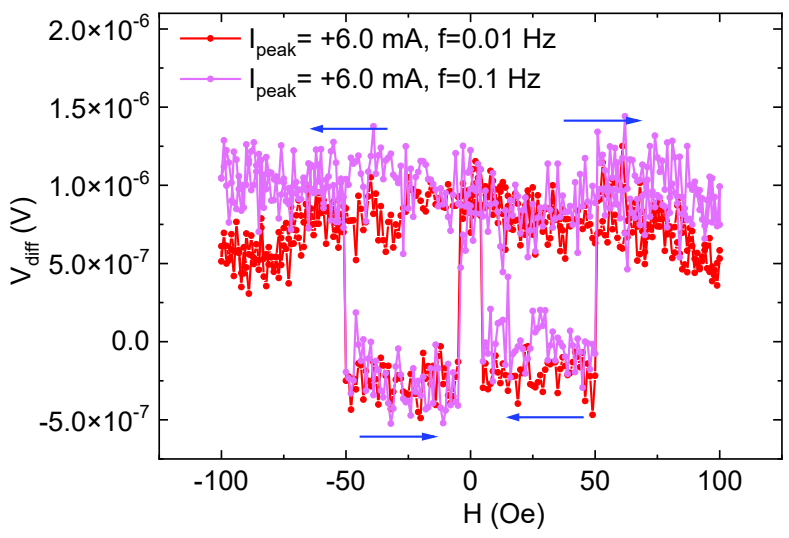

**Fig. S2.** H cycle of differential voltage between the Co electrodes, for applied charge currents in $MoS_2$ $I=I_0+\Delta I \sin(\omega t)$, with $I_{peak}$=+6.0mA and frequencies f=0.01 and 0.1 Hz.

## 2. Current sweeps performed with different protocols

Fig.S3 shows different measurements of differential voltage between the Co electrodes at fixed H, varying the applied sinusoidal current $I=I_0+\Delta I \sin(\omega t)$, always with $\Delta I=|I_0|$. The current $I_0$ was either swept cyclically back and forth (panels S3a and S3b) or swept one-way from zero to a maximum value, either positive or negative, in such a way that no current inversion occurs in each run at fixed field (panels S3c and S3d).

In Fig. S3a, measurements with current cycles are displayed. Note that in this case, not the same hysteresis as in H cycles is expected, due to the different protocol of reversing the sign of the current at each fixed field. It is evident that there is an offset of some curves with respect to others, indicating that an additional non-ohmic voltage is present in some curves. In Fig. S3b, the intercepts of linear fits to data in Fig. S3a are shown. A clear two-level behavior is observed, evidencing this additional non-ohmic voltage. Furthermore, plotting the voltage for the four different branches of the current cycles (inset of Fig. S3b), we observe a similar pattern for H=+100 Oe and H=-100 Oe, and another similar pattern for H=+30 Oe and H=-30 Oe.

In Fig. S3c, the current is swept with $I_{peak}$ from 1 to 6 mA at different fixed field values (+100 Oe, +25 Oe, 0, -100 Oe, -25 Oe and 0, in this sequence), and successively from 1 to -6 mA at the same fixed field values. In this measurement protocol, the current sign is not reversed at fixed field. In this plot, there is the same clear offset of some curves with respect to others, as observed in Fig. S3a. This is better seen in Fig. S3d, where the intercepts of the linear fits are plotted. Two features are clearly observed: a two level behavior and a hysteresis as a function of H, whose clockwise/anticlockwise direction is determined by the sign of the current.

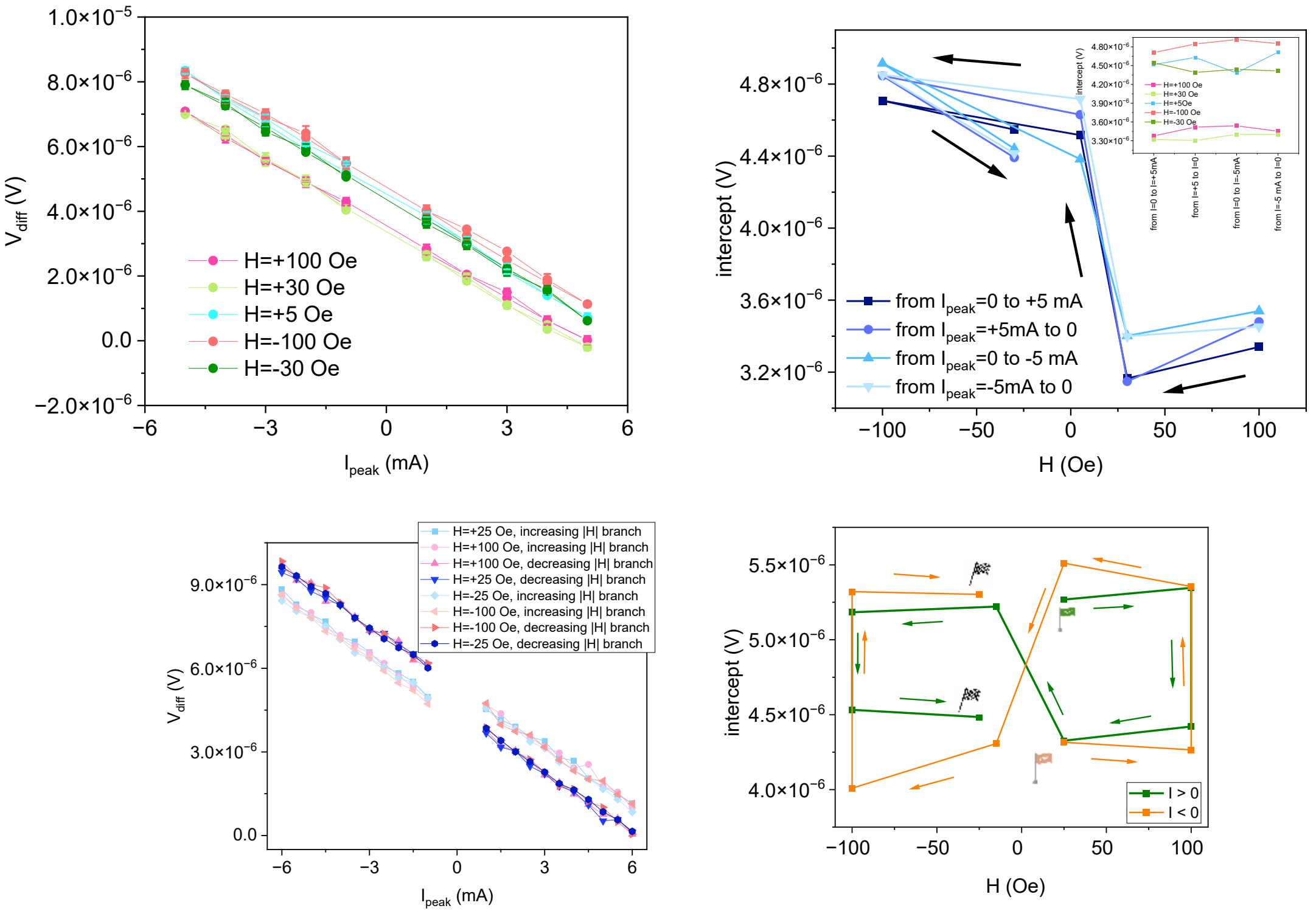

**Figure S3:** a) Differential voltage between Co electrodes, measured in cycles of applied current $I=I_0+\Delta I \sin(\omega t)$, with $I_{peak}$ ($\Delta I=|I_0|$, $I_{peak}=2xI_0$) varied as +1 mA → +5 mA → -1 mA → -5 mA → 0 at fixed H values +100 Oe, +30 Oe, +5 Oe, -100 Oe and -30 Oe, applied in this order. The straight lines are linear fits of data. b) The intercepts of the linear fits of data in panel a) are plotted as a function of the applied field. The arrows indicate the chronological order of current sweeps and the inset shows the same intercept values plotted as a function of the successive branches of the current sweeps.

c) Differential voltage between Co electrodes, measured in sweeps of applied current $I=I_0+\Delta I \sin(\omega t)$, with positive current $I_{peak}$ ($\Delta I=|I_0|$, $I_{peak}=2xI_0$) increasing from +0.5 mA to +6 mA in steps of 0.5 mA, carried out at different fixed field values (100 Oe, 25 Oe, 0, -100 Oe, -25 Oe and 0, in this sequence) and then with negative current $I_{peak}$ ($\Delta I=||I_0||$, $I_{peak}=2xI_0$) varying from -0.5 mA to -6 mA in steps of -0.5 mA, at the same field values. d) Intercepts of the linear fit of panel c) are plotted as a function of the applied field. The arrows indicate the chronological order of current sweeps and the flags indicate the starting and ending points of the measurements for positive and negative current.

## 3. Magnetotransport characterization of the $MoS_2$ flake

Transport properties of the $MoS_2$ flake were characterized. The room temperature resistivity, carrier density and mobility are $3.5x10^{-4}$ $\Omega m$, ~$10^{18}$ $cm^{-3}$ and ~150 $cm^2/(Vs)$, respectively. The resistivity has a semiconducting temperature behaviour with thermal activation energy ~80 meV.

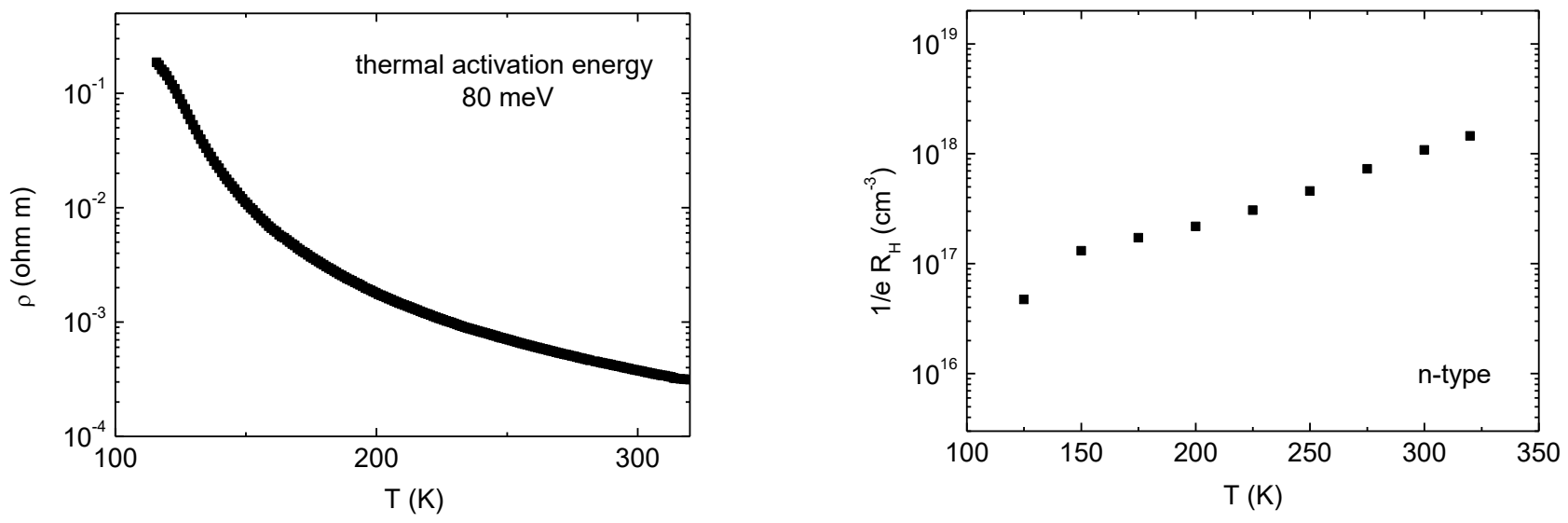

**Fig. S4.** Left panel: temperature dependence of $MoS_2$ resistivity, exhibiting semiconducting behaviour. Right panel: carrier concentration measured by Hall effect.

## 4. Current-voltage characteristics of the $MoS_2$/h-BN/Co heterostructure

In Fig. S5 we show the current-voltage characteristics of the $MoS_2$/h-BN/Co heterostructure. The h-BN resistive spacer exhibits ohmic behavior, with a corresponding resistance of 640 k$\Omega$ over a cross section of 10 $\mu$m x 10 $\mu$m, indicative of a defective h-BN.

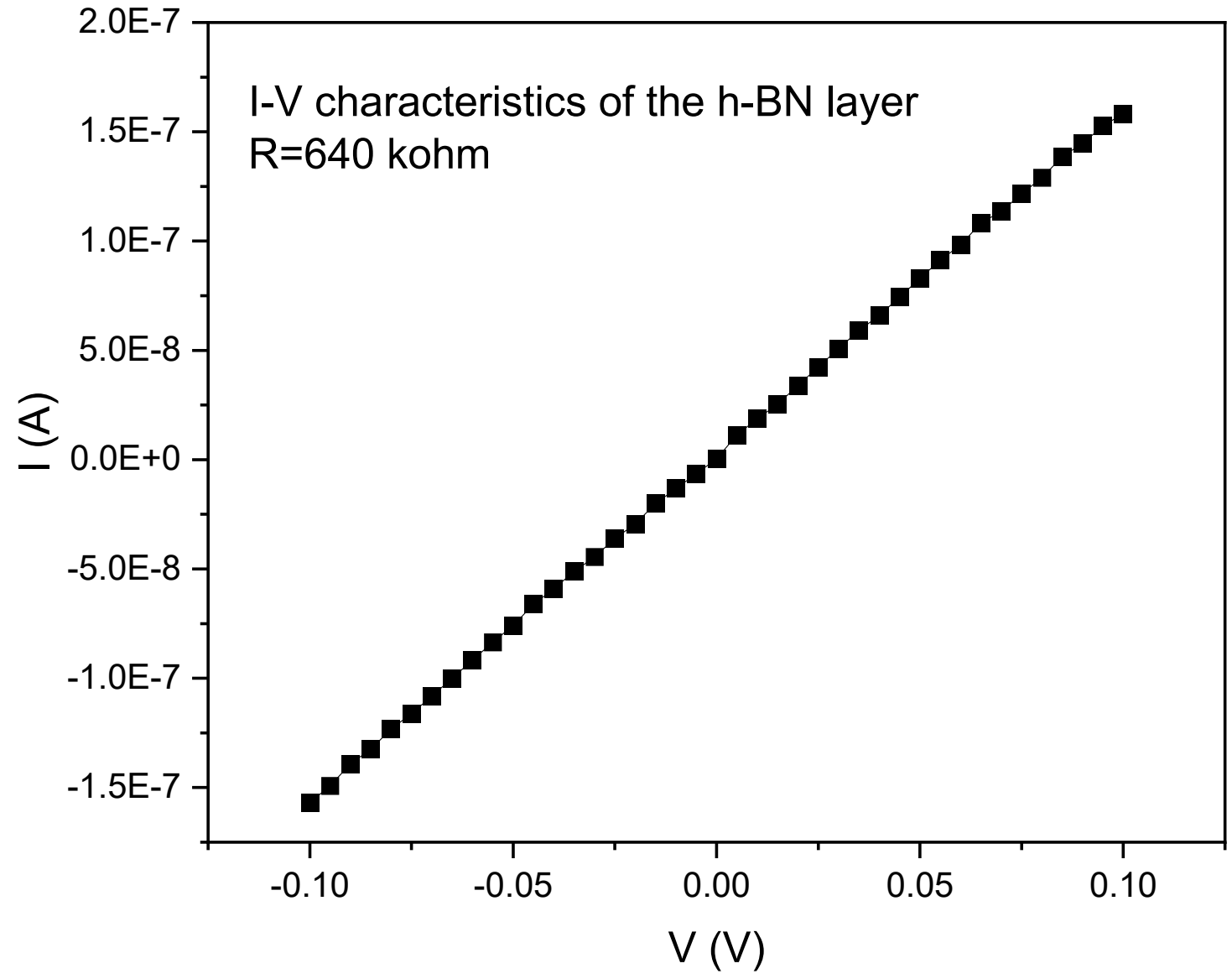

**Fig. S5.** Current-voltage characteristics of the h-BN layer.

## 5. Calculation of SHE and OHE in bulk $MoS_2$

The Berry curvature required for the calculation of SHC and OHC was obtained through Wannier interpolation of the band structure. In Fig. S6, we report the comparison between the *ab initio* (VASP) band dispersion and its fit obtained via Wannierization. The basis used in the Wannierization consists of 44 orbitals: 5 Mo-4d orbitals for each Mo atom and 3 S-3p orbitals for each S atom. The energy window adopted for the Wannierization ranges from −7 to +5 eV. This window contains 44 bands and is separated from the higher excited states by a gap of about 1 eV.

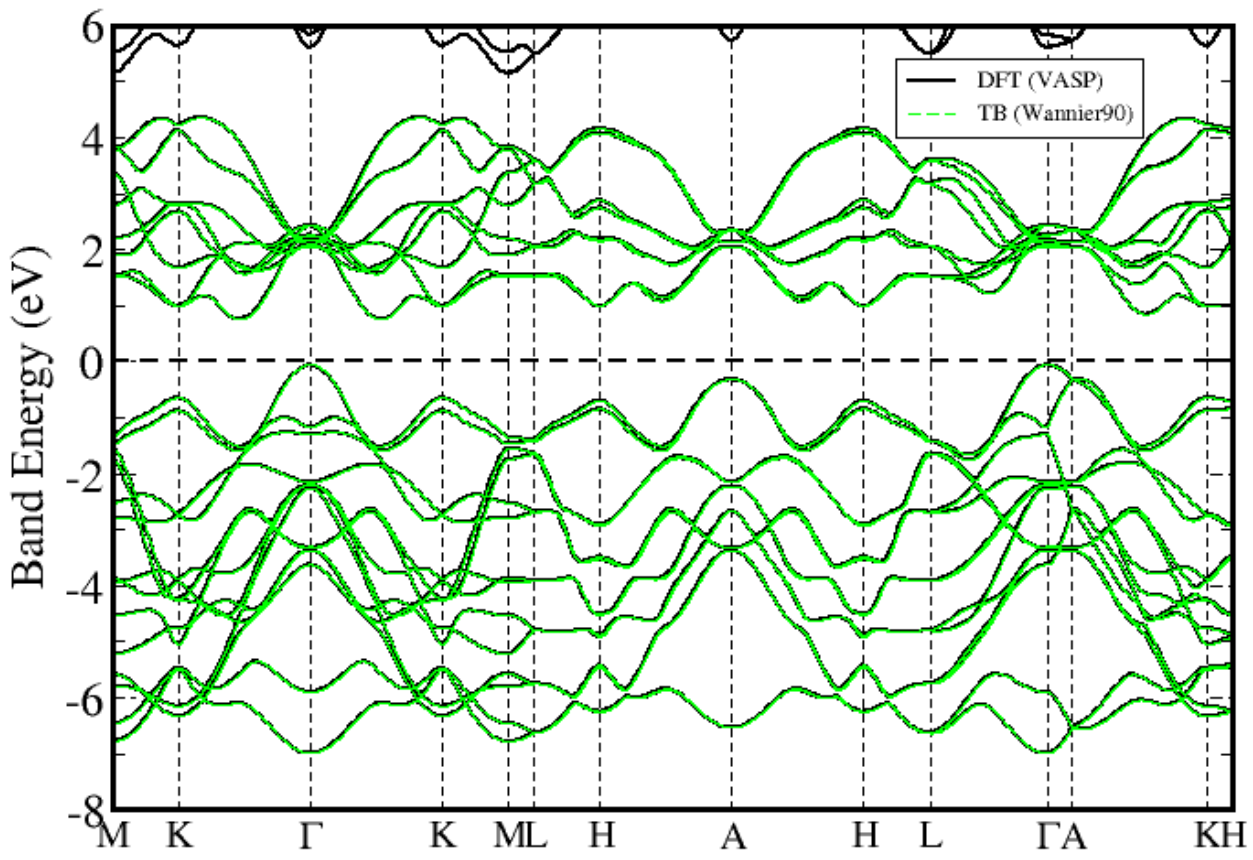

**Fig. S6.** Comparison of the $MoS_2$ band structure obtained from *ab initio* VASP calculations (black solid lines) with the corresponding Wannier-interpolated bands from Wannier90 (green dashed lines).

In this work we were primary interested in the spin and orbital currents flowing along the z-direction (c-axis) and polarized in-plane (y-direction) induced by an electric field with an in-plane direction (x-direction). For completeness and with the same procedure we computed also the other components of the Hall conductivity tensor. Overall, there are 27 components of this tensor. Excluding components having field/current parallel to polarization direction or current parallel to field, there remains 6 components. Exchanging the field with the current direction will change the sign of the conductivity, therefore of the 6 components only three are mutually independent: $\sigma^{z}_{xy}$, $\sigma^{y}_{zx}$ and $\sigma^{x}_{yz}$. In Fig. S7 we show the calculated values for the SHC and OHC in bulk $MoS_2$. We see clearly that SHC vanishes inside the gap, while OHC has a constant finite value. It is also evident that OHC is much higher than the values reached by SHC outside the gap, therefore the orbital current is the dominant source of angular momentum flow inside bulk $MoS_2$.

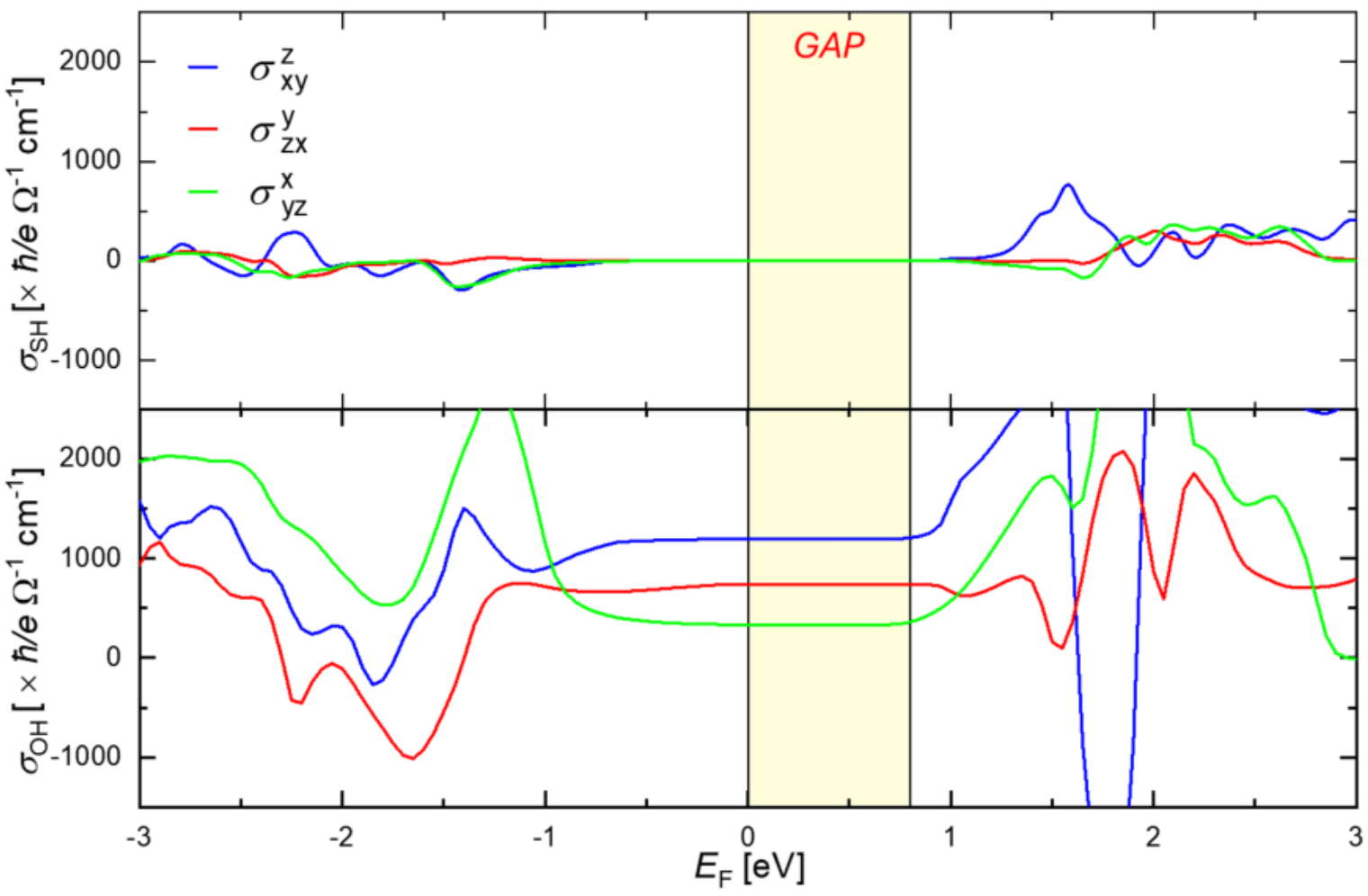

**Fig. S7.** Upper (lower) panel: spin (orbital) Hall conductivity of bulk $MoS_2$. The z-direction is parallel to the c-axis of the hexagonal structure. The x and y-direction are in-plane directions.

## 6. Finite element simulations of current distribution in the $MoS_2$/h-BN/Co heterostructure

We carried out finite element simulation, considering the real device geometry and the measured resistivities, to calculate the current density flowing in each layer, in particular the current shunted along the Co electrodes. In Fig. S8 are some surface plots obtained from the simulation, assuming an applied current I=0.006 A, equal to the maximum current that we applied to our device. It turns out that the current density in the Co layers reaches $2.2 \times 10^6$ A/m$^2$ in the larger Co electrode and $1.0 \times 10^6$ A/m$^2$ in the smaller Co electrode. On the other hand, the current density in the $MoS_2$ layer reaches $8.9 \times 10^8$ A/m$^2$, hence 400 times larger than in the larger Co electrode and 890 times larger than in the smaller Co electrode. These small current density values in Co rule out any role of anisotropic magnetoresistance of Co electrodes in the voltage signal $V_{diff}$ (reading process), as well as any role of self-torque in Co electrodes in the magnetization switching (writing process). We note that the slight asymmetry of current between the two Co electrodes may concur with the tilting in determining their magnetic switching at different applied currents/fields.

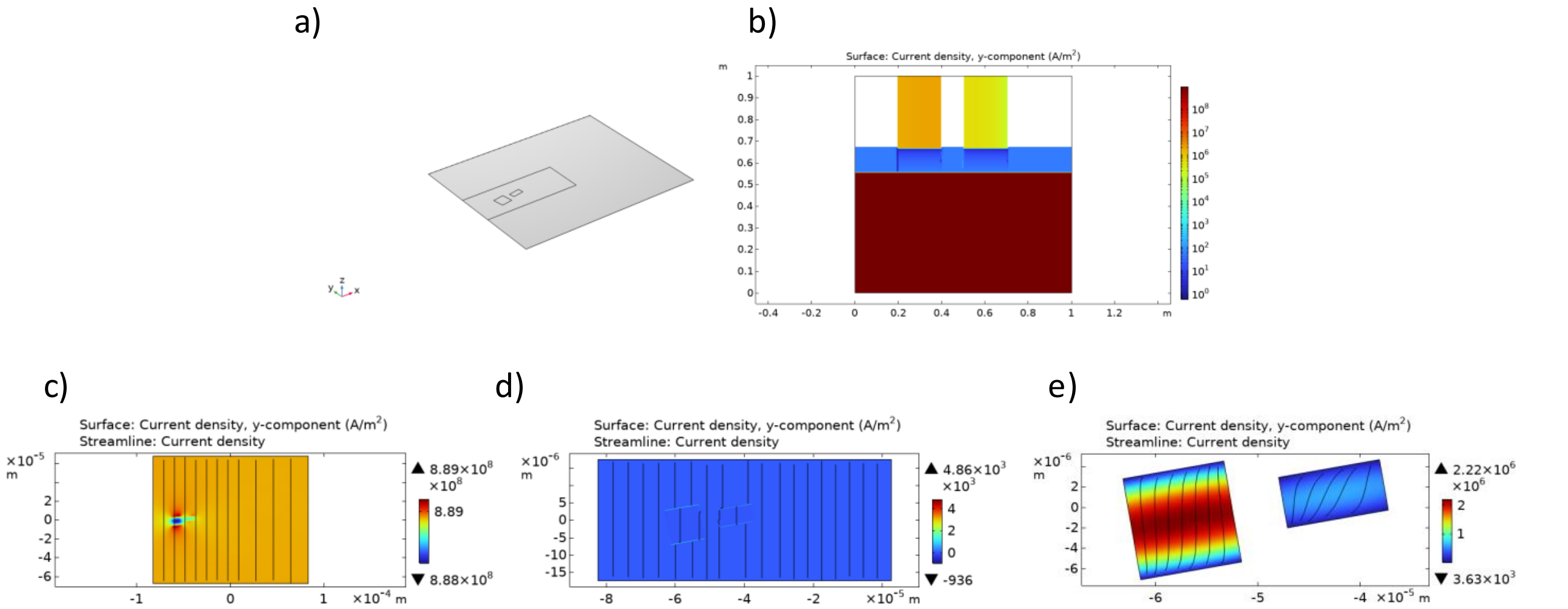

**Fig. S8.** Panels from left to right: a) device geometry and orientation of xyz axes; b) normal current density ($J_y$ component) across a xz cross section, in logarithmic color scale; c) $J_y$ component and uniformly-spaced streamlines of the current density across the mid-plane xy of the $MoS_2$ layer; d) $J_y$ component and uniformly-spaced streamlines of the current density across the mid-plane xy of the h-BN resistive spacer layer; e) $J_y$ component and uniformly-spaced streamlines of the current density across the mid-plane xy of the Co layers. In panels c) to e), color scale legends and maximum and minimum values are indicated on the right-hand side.

## 7. Detailed description of our model of the evolution of micromagnetic configurations in '0' and '1' states in H cycles and in current sweeps.

According to our interpretative picture, represented in Fig. 6 of the main manuscript, three main factors come into play in determining the reading and writing processes on "0" and "1" states of our device:

(1) **Non collinear directions of injected magnetic moments and Co magnetization**. The longitudinal charge current applied to the underlying $MoS_2$ flake generates a vertical current of orbital moments that reaches the Co electrodes. Due to the irregular shape of the $MoS_2$ flake and the peripheral position of the Co electrodes, the direction of charge current flowing beneath the Co electrodes follows the flake edge. Such local current tilting generates magnetic moments that are tilted by an angle $\alpha$ of with respect to the direction of the applied field and of the Co easy magnetization axis (or by an angle ($\pi$-$\alpha$), depending on the sign of the current and on the sign of the field). Additionally, due to the current that follows the flake edge, there could be also a local gradient of charge current density that induces a gradient of transverse spin current, and this could contribute to the magnetization dynamics even in zero field.

(2) **Polydomain nature of Co and role of domain size**. The magnetization of Co electrodes is polydomain with vanishing coercivity, and the domains get increasingly larger as the external field is increased from 0 to 100 Oe. When the domains are fragmented, they rotate reversibly in and out of the magnetic easy axis, under the AC torque generated by the injection of magnetic moments from the underlying $MoS_2$, but when larger domains are formed, they remain fixed either along the direction of the external field or along the direction of the injected magnetic moments, depending on the mutual magnitudes of the related torques and on the magnitude of the applied field with respect to the coercive fields of the enlarged domains of the two different electrodes, 5 Oe and 50 Oe, respectively (see upper left inset of Fig. 6 in the main manuscript).

(3) **Spin-orbit torque and reversible/irreversible switching of Co domains**. The current-generated magnetic moments injected at an angle $\alpha$ in the Co electrodes exert a torque on the magnetic domains. In the ascending branch of the field sweep from zero to +100 Oe, the domains are fragmented and rotate reversibly, following the AC excitation of the injected current and the competition of the torques due to current and field. With increasing field, they become larger and larger, yet they continue to rotate reversibly with the AC current, as long as the field and current are large enough. In the descending branch of the field sweep from +100 Oe to zero, these enlarged domains remain eventually irreversibly oriented by an angle $\alpha$ (for positive charge current), as the field intensity drops below the coercive field of each electrode. This happens firstly, around 50 Oe, for the larger electrode, having lower shape anisotropy, and later, around 5 Oe, for the smaller electrode, having larger shape anisotropy. Hence, for fields between 50 Oe and 5 Oe of the descending branch, the differential voltage between the Co electrodes sets to a different value, corresponding to an average mutual misalignment of Co magnetizations. For fields between 5 Oe and zero, both electrodes are switched irreversibly by the injected magnetic moments to an angle $\alpha$ with respect to the field, and the differential voltage gets back to the initial level representing collinear magnetizations of the two electrodes. When the field changes in sign, the domains are fragmented and reversed. In the negative-H part of the cycle, identically symmetrical mechanisms occur, with the only difference that the angle between the injected magnetic moments and the local Co magnetization is ($\pi$-$\alpha$) instead of $\alpha$. Therefore, the misalignment between the irreversibly switched and the not-yet-irreversibly-switched Co electrodes is larger, and the resulting level of the differential voltage in this non-collinear state between -50 Oe and -5 Oe is further apart from the collinear state voltage level. If the entire H cycle is performed with the opposite polarity of the current applied to $MoS_2$, the shape of the cycle is mirrored with respect to the zero field axis, as the misalignment angle is ($\pi$-$\alpha$) in the H-positive branch and $\alpha$ in the H-negative branch, resulting in swapped values of the non-collinear sub-levels of the differential voltage. We conjecture that the sharpness of the voltage steps suggests monodomain switching at $\pm$5 Oe and $\pm$50 Oe.

Our tentative micromagnetic model sketched in Fig. 6 of the main manuscript, applies not only to the H cycles, but also to the current cycles (Fig.s 3 and 4 of the main manuscript). From the experimental measurements in Fig. 3, we assume that not only the increasing field, but also the increasing current is responsible for the enlargement of domains in the first current sweep, so that while in the first current sweep the domains rotate reversibly under the opposed current- and field-induced torques, in the successive identical current sweeps the domains switch irreversibly to the direction of the injected magnetic moments. As the current is reversed to a value of opposite sign that is close to the threshold value, the domains get fragmented again and a new current sweep that makes the domains grow is necessary to enable the irreversible switching in the successive sweep. As comes out from the summary of switching sequences displayed in Fig. 3c of the main manuscript, in this regime of current-driven domain growth, zero or low fields (H < 50 Oe) do not influence the irreversible current-induced switching, which is only determined by the sign of the applied current. On the other hand, at higher fields H > 50 Oe, the irreversible current-induced switching is determined by the sign of the applied current, as well as by the sign of the field. In order to better explain the switching in zero field, in Fig. 7 of the main manuscript we sketch the evolution of magnetic domains in the Co electrodes in a current cycle as that of Fig. 3a of the main manuscript. The two different output voltage $V_{diff}$ states correspond to the following possible magnetic configurations: in the '0' state, the magnetizations of the Co electrodes are either both fragmented into small domains with spread orientations or both with parallel monodomains; in the '1' state, the magnetizations of the Co electrodes are non-collinear, forming an angle either $\alpha$ or ($\pi$-$\alpha$).

Let's start with the '0' state in zero field. A current with increasing magnitude injects magnetic moments into the Co electrodes and enlarges the domains therein, aligning them along the magnetic moments $\mu$, due to the damping-like torque $\tau_\mu \sim M \times (\mu \times M)$. As long as the current exceeds a threshold, the magnetization of the larger square electrode with no shape anisotropy gets rotated irreversibly along the injected magnetic moments, while the magnetization of the smaller rectangular electrode aligns along its easy axis. The '1' state is thus obtained this way by applying a positive or negative current with above-threshold intensity. Two different magnetic configurations can be established, depending on the positive or negative sign of the applied above-threshold current, having an angle either $\alpha$ or ($\pi$-$\alpha$) between the electrode magnetizations, respectively, and both corresponding to the same voltage state '1'. These two states, '1a' and '1b', are depicted in Fig. 7 of the main manuscript. To switch back to the '0' state, a current of opposite sign and close-to-threshold of above-threshold intensity must be applied, which fragments the domains. Hence the sign of the current to be applied to switch back from the '1'state to the '0' state depends on which of the two magnetic configurations, either '1a' or '1b', is presently realized. For this reason, an initialization magnetization protocol must be used, with application of above-threshold applied field and current, to fix a given initial configuration of magnetizations, so that the polarity of the current that must be applied to switch deterministically the state is established.

On the other hand, in the presence of a magnetic field applied along the easy axis of the smaller Co electrode, whose intensity exceeds the coercive field ~50 Oe, the symmetry is broken and the initial state of the electrodes magnetization is fixed, so that also the sign of the current to be applied to switch from the '1' state to the '0' state is known. This current sign is opposite for opposite signs of the applied field, as observed experimentally in the upper panel of Fig. 3c of the main manuscript.

We think that, due to the low magnetic anisotropy of the Co electrodes, it is plausible to assume that reversing current or field drives the Co electrodes in a polydomain state, corresponding to the '0' voltage state. This could enable the current-induced deterministic switching mechanism, without invoking any role of a current gradient in breaking the symmetry [1,2], which on the other hand could be possibly present.

## 8. Anisotropic magnetoresistance measurements in Cobalt films

In Fig. S9 we show measurements of anisotropic magnetoresistance measured in a Cobalt film patterned as a Hall bar with 10 μm wide channel, in an applied in-plane field of at 200 Oe and 500 Oe. The measured resistance exhibits four-fold symmetry and it changes by 0.4 % by varying the angle ϕ between the applied current field.

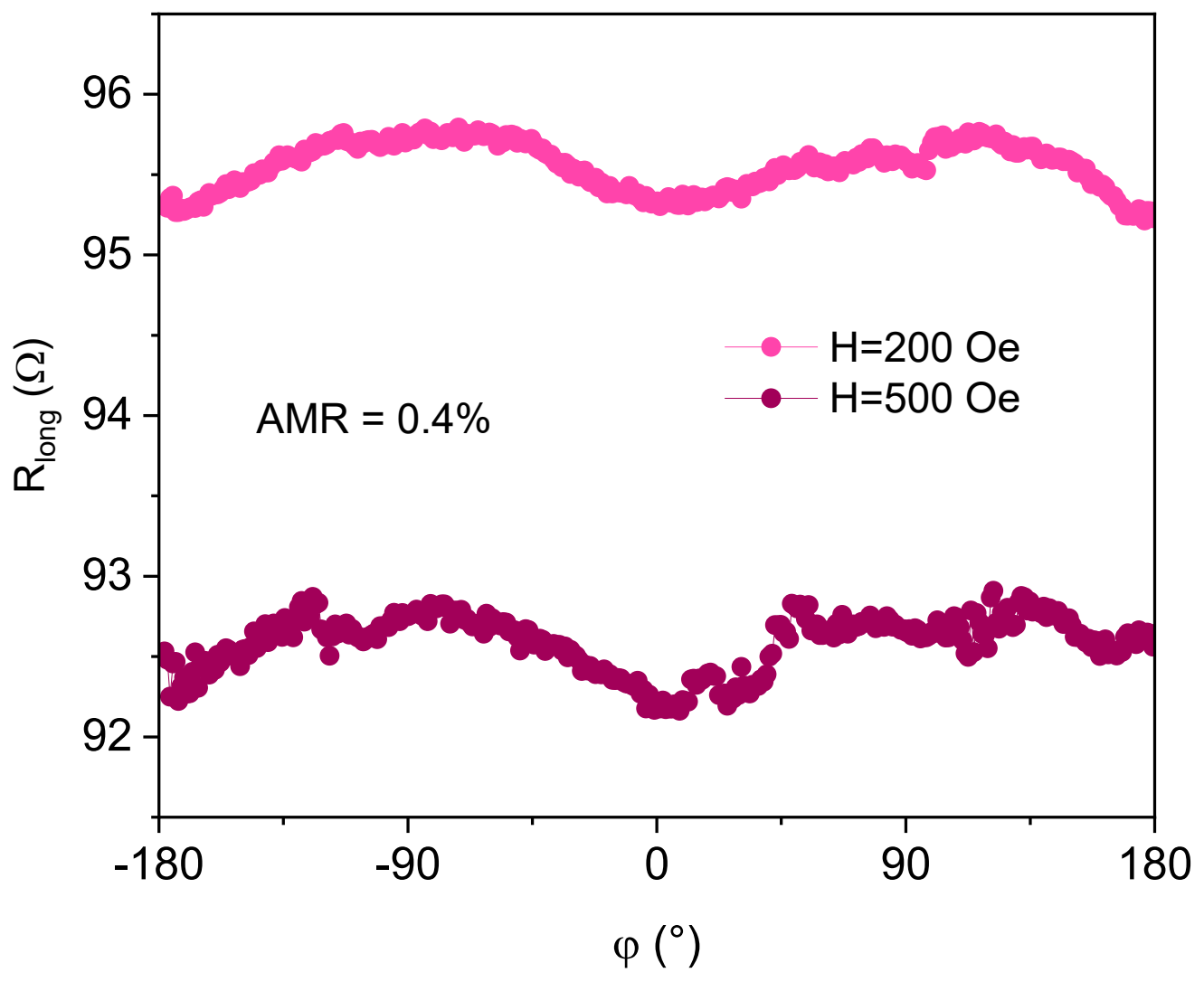

**Fig. S9.** anisotropic magnetoresistance measured in a Cobalt Hall bar.